\documentclass[fleqn,usenatbib]{mnras}

\usepackage{newtxtext,newtxmath}

\usepackage{caption}
\usepackage{subcaption}
\usepackage[T1]{fontenc}
\usepackage{multirow}
\usepackage{array}
\usepackage{soul}
\usepackage{xcolor}
\usepackage[noabbrev]{cleveref}

\DeclareRobustCommand{\VAN}[3]{#2}
\let\VANthebibliography\thebibliography
\def\thebibliography{\DeclareRobustCommand{\VAN}[3]{##3}\VANthebibliography}

\usepackage{graphicx}	
\usepackage{amsmath}	

\def\prd{{\em Phys. Rev. }{\bf D }}
\def\prl{{\em Phys. Rev. }{\bf L }}

\newcommand{\lya}{Ly$\alpha$}
\newcommand{\lyaf}{Ly$\alpha$ forest}
\newcommand{\lyb}{Ly$\beta$}
\newcommand{\lyalya}{Ly$\alpha\times$Ly$\alpha$}
\newcommand{\lyaqso}{Ly$\alpha\times$QSO}

\newcommand{\lyalyalyalya}{Ly$\alpha$(Ly$\alpha$)$\times$Ly$\alpha$(Ly$\alpha$)}
\newcommand{\lyalyalyalyb}{Ly$\alpha$(Ly$\alpha$)$\times$Ly$\alpha$(Ly$\beta$)}
\newcommand{\lyalyaqso}{Ly$\alpha$(Ly$\alpha$)$\times$QSO}
\newcommand{\lyalybqso}{Ly$\alpha$(Ly$\beta$)$\times$QSO}

\newcommand{\hMpc}{\ $h^{-1}\text{Mpc}$}

\newcommand{\phip}{{$\phi_\mathrm{p}$}}
\newcommand{\phis}{{$\phi_\mathrm{s}$}}
\newcommand{\phif}{{$\phi_f$}}
\newcommand{\alphap}{{$\alpha_\mathrm{p}$}}
\newcommand{\alphas}{{$\alpha_\mathrm{s}$}}

\newcommand{\colore}{\texttt{CoLoRe}}
\newcommand{\lyacolore}{\texttt{LyaCoLoRe}}

\title[AP from \lya\ correlations in synthetic data]{The Alcock-Paczy\'nski effect from Lyman-$\alpha$ forest correlations: Analysis validation with synthetic data}

\author[A. Cuceu et al.]{Andrei Cuceu,$^{1,2,3,4}$\thanks{E-mail: cuceu.1@osu.edu}
Andreu Font-Ribera,$^{5,4}$
Paul Martini,$^{2,1}$
Benjamin Joachimi,$^{4}$
Seshadri Nadathur,$^{6}$
\newauthor
James Rich,$^{7}$
Alma~X.~Gonz\'alez-Morales,$^{8}$
H\'elion~du~Mas~des~Bourboux$^{9}$
and James~Farr$^{4}$
\\
$^{1}$Center for Cosmology and Astro-Particle Physics, The Ohio State University, Columbus, Ohio 43210, USA\\
$^{2}$Department of Astronomy, The Ohio State University, Columbus, Ohio 43210, USA\\
$^{3}$Department of Physics, The Ohio State University, Columbus, Ohio 43210, USA\\
$^{4}$Department of Physics and Astronomy, University College London, Gower Street, London WC1E 6BT, United Kingdom\\
$^{5}$Institut de Física d’Altes Energies, The Barcelona Institute of Science and Technology, Campus UAB, 08193 Bellaterra (Barcelona), Spain\\
$^{6}$Institute of Cosmology and Gravitation, University of Portsmouth, Burnaby Road, Portsmouth, PO1 3FX, United Kingdom\\
$^{7}$IRFU, CEA, Universit\'e Paris-Saclay, F91191 Gif-sur-Yvette, France\\
$^{8}$Departamento de F\'isica, DCI, Campus Le\'on, Universidad de Guanajuato, 37150, Le\'on, Guanajuato, M\'exico\\
$^{9}$Department of Physics and Astronomy, University of Utah, 115 S 1400 E, Salt Lake City, UT 84112, USA
}

\date{Accepted XXX. Received YYY; in original form ZZZ}

\pubyear{2022}

\begin{document}
\label{firstpage}
\pagerange{\pageref{firstpage}--\pageref{lastpage}}
\maketitle

\begin{abstract}
The three-dimensional distribution of the \lyaf\ has been extensively used to constrain cosmology through measurements of the baryon acoustic oscillations (BAO) scale. However, more cosmological information could be extracted from the full shapes of the \lyaf\ correlations through the Alcock-Paczy\'nski (AP) effect. In this work, we prepare for a cosmological analysis of the full shape of the \lyaf\ correlations by studying synthetic data of the extended Baryon Oscillation Spectroscopic Survey (eBOSS). We use a set of one hundred eBOSS synthetic data sets in order to validate such an analysis. These mocks undergo the same analysis process as the real data. We perform a full-shape analysis on the mean of the correlation functions measured from the one hundred eBOSS realizations, and find that our model of the \lya\ correlations performs well on current data sets. We show that we are able to obtain an unbiased full-shape measurement of $D_M/D_H(z_\mathrm{eff})$, where $D_M$ is the transverse comoving distance, $D_H$ is the Hubble distance, and $z_\mathrm{eff}$ is the effective redshift of the measurement. We test the fit over a range of scales, and decide to use a minimum separation of $r_\mathrm{min}=25$\hMpc. We also study and discuss the impact of the main contaminants affecting \lyaf\ correlations, and give recommendations on how to perform such analysis with real data. While the final eBOSS \lya\ BAO analysis measured $D_M/D_H(z_\mathrm{eff}=2.33)$ with $4\%$ statistical precision, a full-shape fit of the same correlations could provide a $\sim2\%$ measurement.
\end{abstract}




\section{Introduction}
\label{sec:intro}

Over the last few decades, probes of the large-scale structure (LSS) of the Universe have become one of the main tools for studying its expansion history and the properties of its constituents. Measuring the scale of the baryon acoustic oscillations (BAO) feature is currently the most widely used method for extracting cosmological information from LSS \citep[e.g.][]{Eisenstein:2005,Cole:2005,Beutler:2011,Ross:2015,Alam:2017,Abbott:2019,Alam:2021,Abbott:2022}. BAO produces distinct features in the two-point statistics of LSS probes, which allows us to disentangle its signal from the other cosmological, astrophysical, and instrumental effects present. In BAO analyses, other cosmological information is effectively ignored by marginalizing over the broadband of the two-point statistics.

A common way of measuring this extra cosmological information is to instead model the full shape of the correlation function or power spectrum in order to measure the Alcock-Paczy\'nski (AP) effect and redshift space distortions \citep[RSD; e.g.][]{Blake:2011,Reid:2012,Beutler:2012,Samushia:2014}. However, such analyses are more difficult than BAO analyses, because accurately modelling these effects requires a good understanding of the astrophysical and instrumental contaminants present in the broadband of two-point statistics. Furthermore, linear theory approaches that may work well on large scales start to break down due to non-linear growth on smaller scales. These difficulties have been successfully addressed in the case of two-point statistics of discrete tracers, leading to numerous full-shape analyses in the literature \citep[e.g.][]{Blake:2011,Reid:2012,Beutler:2012,Samushia:2014,Satpathy:2017,Beutler:2017,GilMarin:2020,Bautista:2021,Hou:2021,Neveux:2020}. On the other hand, in the case of the three-dimensional (3D) distribution of the Lyman-$\alpha$ (\lya) forest, this type of measurement has not yet been performed \citep{Cuceu:2021}.

The \lyaf\ is made up of absorption lines blueward of the \lya\ emission peak in spectra of high-redshift quasars \citep[e.g.][]{Lynds:1971,Rauch:1998}. This absorption is caused by neutral hydrogen in the intergalactic medium between the quasar and us. Due to the expansion of the Universe, the \lya\ line is progressively redshifted as the photons travel towards us. This means the distribution of \lya\ absorption can be used to study the structure and evolution of the Universe. The forest has long been used as a tool for cosmology through its one-dimensional\footnote{The one dimension here is along the line-of-sight.} two-point statistics \citep[e.g.][]{McDonald:2006,Seljak:2005,Viel:2010,Palanque:2015,Palanque:2020}. With the advent of large spectroscopic surveys such as the Baryon Oscillation Spectroscopic Survey (BOSS) and its successor, extended BOSS (eBOSS), the 3D distribution of the \lyaf\ has also become a useful source of cosmological information \citep{Busca:2013,Slosar:2013,Kirkby:2013,FontRibera:2014,Delubac:2015,Bautista:2017,duMasdesBourboux:2017,deSainteAgathe:2019,Blomqvist:2019,duMasdesBourboux:2020}. However, as mentioned above, it has only been used to measure BAO.

The BOSS and eBOSS \lya\ cosmological analyses have progressively improved our understanding of the different processes that contribute to the auto-correlation function of \lya\ transmitted flux, and its cross-correlation with the quasar distribution. The first physical model of the two correlations was introduced with the BOSS data release (DR) 12 analyses \citep{Bautista:2017,duMasdesBourboux:2017}. The main effects that had to be understood were the distortion caused by fitting the quasar continuum (see \citealt{Bautista:2017,Perez-Rafols:2018} for the modern treatment and \citealt{Slosar:2011,Font-Ribera:2012a,Blomqvist:2015} for further discussion of the effect), the contamination due to metal absorption in the forest \citep{Pieri:2010,Pieri:2014}, and the contamination due to high column density (HCD) absorbers \citep{Font-Ribera:2012a,Font-Ribera:2012b}. Starting with the eBOSS DR14 analyses, \lya\ absorption in the \lyb\ region (blueward of the \lyb\ peak) was also used, first through its correlation with \lya\ absorption in the \lya\ region \citep{deSainteAgathe:2019}, and then through its cross-correlation with quasars \citep{Blomqvist:2019}. The model of the \lya\ correlation functions has been progressively improved through these analyses, and we now may be in a position where we can perform a full-shape analysis. The goal of this article is to test this. In particular, we want to study the performance of this model when it comes to fitting the full-shape of the eBOSS DR16 \lyaf\ correlation functions.

As mentioned above, a common way of extracting cosmological information from the full shape of two-point statistics is to measure the AP effect and RSD. The AP effect arises due to the fiducial cosmology used to transform redshifts and angles into comoving coordinates. Any differences between this fiducial cosmology and the true cosmology will introduce an extra anisotropy in the auto- and cross-correlations \citep{Alcock:1979}. Assuming we can accurately model other sources of anisotropy (e.g. RSD, HCDs, metals, distortion due to continuum fitting), the remaining anisotropy can be used to constrain the background cosmological model. This method is already used in analyses of the BAO, but these only use a small range of scales around the BAO peak. \cite{Cuceu:2021} have shown that measuring the AP effect using the full shape is a promising avenue for extracting more cosmological information from the \lyaf. RSD measurements themselves are also generally used to measure the growth rate of cosmic structure, but for the \lya\ auto-correlation this is degenerate with an unknown RSD bias parameter \citep{McDonald:2003,Slosar:2011,Givans:2020,Chen:2021}. \cite{Cuceu:2021} did show that it is possible to measure the growth rate from the quasar RSD parameter in a joint analysis of the \lya\ auto-correlation and its cross-correlation with quasars. However, this would not be a significant measurement with eBOSS. Therefore, in this work we only focus on the AP information, and leave RSD measurements for future work using larger surveys such as the Dark Energy Spectroscopic Instrument (\citealt{DESI:2016}).

Our objective in this work is to analyse the full shape of the \lyaf\ correlation functions in simulated data in order to understand how well current models perform relative to the expected precision of the eBOSS DR16 data. This analysis is meant as a first step towards a full-shape analysis of the eBOSS \lya\ 3D correlation functions. The full-shape measurement using real eBOSS data is presented in a separate article \citep{Cuceu:2022}.

This article is structured as follows. In \Cref{sec:framework}, we introduce our framework, which includes a description of the mock data sets, the analysis process for computing the 3D correlation functions, and the modelling of the correlations. We show the results of our analysis in \Cref{sec:results}, and discuss their implications for a full-shape analysis using real data in \Cref{sec:discussion}. We summarize and conclude in \Cref{sec:conclusion}.


\section{Framework}
\label{sec:framework}

We begin with an overview of the framework we use to validate a full-shape analysis of the \lyaf\ 3D correlation functions. We use the set of one hundred eBOSS mock data sets produced by \cite{duMasdesBourboux:2020}, which were created using the method introduced by \cite{Farr:2020}. We describe these mock data sets in \Cref{subsec:mocks}, and the measurement of the transmitted flux correlations in \Cref{subsec:spec_to_corr}. The analysis methodology we use is similar to the one used in past BAO analyses of \lya\ BOSS and eBOSS data. The only differences appear at the step of fitting the correlation functions. Instead of only fitting for the BAO peak position, we fit the full shape of the correlations in order to extract the AP parameter. This is based on the method introduced in \cite{Cuceu:2021}. We describe our modelling framework in \Cref{subsec:model}. 


\subsection{Synthetic data-sets}
\label{subsec:mocks}

In order to test the model of the \lya\ correlation functions, we use the set of one hundred mock data sets introduced in \cite{duMasdesBourboux:2020}. These mocks were produced for the BAO analysis of the \lya\ eBOSS DR16 data set. Each mock is based on a Gaussian random field generated with the \colore\footnote{https://github.com/damonge/CoLoRe} package \citep{Ramirez:2022}. The Gaussian field is transformed into a log-normal density field, which is Poisson sampled based on an input number density and bias in order to obtain a set of quasar sources. \colore\ then computes line-of-sight skewers by interpolating the initial Gaussian field and the radial velocity field\footnote{Computed using the gradient of the Newtonian potential.} from each quasar to the centre of the box.

The skewers generated by \colore\ require significant post-processing in order to turn them into realistic simulations of the \lyaf. This is performed by the \lyacolore\footnote{https://github.com/igmhub/LyaCoLoRe} package \citep{Farr:2020}. In order to create the large boxes needed to simulate an all-sky light-cone to $z=3.7$ ($\sim10$ $h^{-1}\text{Gpc}$ in this case), the \colore\ grid is limited by memory and computational constraints to a resolution of $\sim2.4$\hMpc\ (for the $4906^3$ grid used here). However, one of the contributions to the covariance of 3D correlations (and consequently of BAO uncertainties) is related to the amount of small-scale fluctuations in the \lyaf\ \citep{McDonald:2007}, often characterized by the one-dimensional power spectrum, or $P_{\rm 1D}$. In order to simulate spectra with a realistic $P_{\rm 1D}$, \lyacolore\ adds an extra 1D Gaussian field to each line-of-sight. After that, the field undergoes a log-normal transformation, and the output is used to compute the optical depth field ($\tau$) using the fluctuating Gunn-Peterson approximation \citep[FGPA;][]{Bi:1997,Croft:1998}. Redshift-space distortions (RSD) are introduced by convolving the real-space optical depth field with a kernel based on the peculiar velocity field. Finally, the transmitted flux in redshift-space ($s$) is given by $F(s) = \exp{[-\tau(s)]}$. For more details on this process, and how it is tuned to produce realistic forests, see \cite{Farr:2020}.

Beyond the \lyaf\ transmitted flux fraction, \lyacolore\ also simulates high column density (HCD) absorbers, as well as absorption by other Lyman lines and metals. HCDs are sampled from the Gaussian density field using an input bias and number density based on literature results. After that, each HCD is allocated a column density given a column density distribution from the literature. In contrast, the other absorption lines are produced using a rescaled version of the \lya\ optical depth, which is then mapped to a different observed wavelength based on the rest-frame wavelength of the absorber. In the case of higher Lyman lines, the scaling factors are tuned based on the oscillator strengths of each transition \citep{Irsic:2013,Farr:2020}, with the scaling factor of  Ly$\beta$ being $0.1901$. For metal absorbers, the scaling factors are tuned in order to reproduce the level of contamination in the data \citep{Farr:2020,duMasdesBourboux:2020}. A list of the main metal absorbers along with their relative strength can be found in Table 2 of \citet{Farr:2020}. These contaminants (HCDs, metal lines, and higher-order hydrogen lines) are then added to the simulated \lyaf\ transmitted flux fraction.

Each quasar is assigned a random magnitude using an input quasar luminosity function based on \cite{Palanque:2016}. This is used to generate an unabsorbed continuum for each quasar by adding a set of emission lines on top of a broken power law \citep{duMasdesBourboux:2020}. Random redshift errors are drawn from a Gaussian distribution with a standard deviation $\sigma_z=400 \text{ km s}^{-1}$. The \texttt{specsim} package \citep{Kirkby:2016} is then used to simulate the eBOSS spectral resolution, the pixelization of the spectra, and instrumental noise \citep{duMasdesBourboux:2020}.

While these synthetic data sets were created for BAO studies, in this article we use them to study full-shape analyses. As described above, the major contaminants that are known to affect \lyaf\ correlations are being modelled in these mocks. This will allow us to test the performance of our model, and how full-shape analyses are affected by these contaminants. However, one of the main concerns when attempting to use the full-shape of correlation functions for cosmology is the ability to accurately model the non-linear effects on small scales. While we will be able to test some of these using the mocks described above, some ingredients are missing. These include the fact that the small-scale quasar clustering is overestimated \citep{Youles:2022}, and missing IGM effects such as pressure smoothing and the impact of UV radiation on the ionization fraction. We will discuss these issues and the applicability of our results in more detail in \Cref{sec:discussion}.


\subsection{From spectra to correlation functions}
\label{subsec:spec_to_corr}

Once the simulated eBOSS spectra are produced, they undergo the same analysis process as the real data, in order to measure the 3D auto-correlation of \lya\ transmitted flux, and its cross-correlation with the quasar distribution. This is described in detail in \cite{duMasdesBourboux:2020}. Here, we only give a brief overview of this process, as all the results in this article\footnote{With the exception of those in \Cref{app:fid_cosmo}.} come from the same mock correlation functions computed for the eBOSS \lya\ BAO analysis.

The first step in the analysis process is to compute the flux transmission fluctuation, $\delta_q(\lambda)$, of each quasar $q$, based on the observed flux, $f_q(\lambda)$:
\begin{equation}
    \delta_q(\lambda) = \frac{f_q(\lambda)}{\Bar{F}(\lambda)C_q(\lambda)}-1,
\end{equation}
where $\Bar{F}(\lambda)$ is the global mean transmission, and $C_q(\lambda)$ is the unabsorbed quasar continuum. In general, the true quasar continuum is unknown, and therefore, the data is used to jointly fit for the product $\Bar{F}(\lambda)C_q(\lambda)$. This fit also necessarily includes density modes of the size of the forest and larger, which biases $\delta_q$ towards zero for each forest. Even though pixels from the same forest are not cross-correlated, this will still bias pixel correlations in nearby forests, resulting in a distortion of the correlation functions.

The distortion of the correlation function can be removed by building a projection, $\eta_{ij}^q$, such that:
\begin{equation}
    \sum_j \eta_{ij}^q\delta_q^m(\lambda_j) = \sum_j \eta_{ij}^q\delta_q^t(\lambda_j),
    \label{eq:projection}
\end{equation}
where $\delta_q^m$ and $\delta_q^t$ are the measured and true flux fluctuations, $j$ indexes forest pixels before the projection, and $i$ indexes the projected forest pixels. Therefore, the left-hand side of \Cref{eq:projection} is applied to the measured flux fluctuation field, while the right-hand side is forward modelled in the correlations, as described in \Cref{subsec:model}. A detailed description of this projection and how it is built can be found in \cite{Bautista:2017}, and Appendix B of \cite{Perez-Rafols:2018}.

\begin{table*}
\begin{center}
    \begin{tabular}{|c||c|c|c|}
        \hline
        Type & Name & Tracer 1 & Tracer 2 \\
        \hline \hline
        \multirow{2}{2.2em}{Auto} & \lyalyalyalya & \lya\ flux in the \lya\ region & \lya\ flux in the \lya\ region \\
        & \lyalyalyalyb & \lya\ flux in the \lya\ region & \lya\ flux in the \lyb\ region \\
        \hline
        \multirow{2}{2.2em}{Cross} & \lyalyaqso & \lya\ flux in the \lya\ region & Quasars \\
        & \lyalybqso & \lya\ flux in the \lyb\ region & Quasars \\
        \hline
    \end{tabular}
    \caption{The types and names of the four correlation functions we use in this work, along with the tracers used in each of them. Here, we use "flux" to refer to the transmitted flux fraction. The \lya\ and \lyb\ regions are defined by the rest-frame intervals $\lambda_\mathrm{RF}\in [104, 120]\;$nm (situated between the \lya\ and \lyb\ peaks) and $\lambda_\mathrm{RF}\in [92, 102]\;$nm (situated blueward of the \lyb\ peak), respectively.}
    \label{tab:corr}
\end{center}
\end{table*}

Forests that contain HCDs with column densities $\log N_{\mathrm{H}\textsc{i}}>20.3$\footnote{This is close to the detection threshold of the DLA detector used when analysing real data. \cite{duMasdesBourboux:2020} also tested the DLA detector on the mock spectra in order to validate this approach.} are given special treatment. Firstly, the regions where the HCD reduces the transmission by more than $20\%$ are masked. Secondly, the absorption wings are corrected using a Voigt profile \citep{duMasdesBourboux:2020,Noterdaeme:2012}. HCDs with column densities $\log N_{\mathrm{H}\textsc{i}}<20.3$ remain in the data, and their effect has to be included in the model of the correlations (\Cref{subsec:model}).

The correlation functions were computed on a grid in comoving coordinates. For two tracers, $i$ and $j$, at redshifts $z_i$ and $z_j$, and separated by an angle $\Delta\theta$, the comoving separations along and across the line-of-sight are given by \citep{deSainteAgathe:2019}:
\begin{align}
    r_{||} &= [D_\mathrm{c}(z_i) - D_\mathrm{c}(z_j)] \cos{\frac{\Delta \theta}{2}}, \\
    r_\bot &= [D_\mathrm{M}(z_i) + D_\mathrm{M}(z_j)] \sin{\frac{\Delta \theta}{2}},
\end{align}
where $D_\mathrm{M}$ is the transverse comoving distance, $D_\mathrm{c}(z) = c \int_0^z dz'/H(z')$ is the radial comoving distance, with $H(z)$ as the Hubble parameter and $c$ as the speed of light. In this work, we will also refer to the $(r,\mu)$ parametrization, defined as $r^2 = r_{||}^2 + r_\bot^2$ and $\mu = r_{||}/r$. \cite{duMasdesBourboux:2020} used a fiducial cosmological model\footnote{The fiducial cosmology is based on fits to cosmic microwave background anisotropy data from the Planck satellite \cite{Planck:2016}; see Table 2 of \cite{duMasdesBourboux:2020}.} to compute the two distances, which will result in an extra anisotropy in the measured correlations if it differs from the true cosmology. This is the Alcock-Paczy\'nski effect \citep{Alcock:1979}.

The correlation function is first computed independently in each HEALPix\footnote{https://healpix.sourceforge.io} pixel \citep{Gorski:2005}. For the eBOSS footprint, there are about $880$ pixels (\texttt{nside} $=16$), with each covering $3.7\times3.7=13.4 \;\mathrm{deg}^2$. This corresponds to a $250\times250 \;(h^{-1}\mathrm{Mpc})^2$ patch at $z_\mathrm{eff}=2.33$. The population of correlations can then be used to compute the mean and covariance of the correlation function of the entire survey. When computing the correlation function in one HEALPix pixel, pairs with forests in neighbouring pixels are still counted. However, given that we are most sensitive to small scales when measuring AP, we assume that the correlation function measurements in each HEALPix pixel are independent for the purposes of computing the covariance matrix. This method of computing the covariance matrix of \lyaf\ correlation functions has been validated against other methods by \cite{duMasdesBourboux:2020}.

The process for computing the cross-correlation with the quasar distribution, and its covariance matrix, is similar to the one used for the auto-correlation. However, in this case we also distinguish between \lya\ flux in front of a quasar, which is assigned negative $r_{||}$, and \lya\ flux behind a quasar, which is assigned positive $r_{||}$. This is because the cross-correlation is not symmetric under permutations of the two tracers.

Four types of correlation functions were computed by \cite{duMasdesBourboux:2020} for the eBOSS DR16 \lya\ BAO analysis. The first two are the auto-correlation of the \lya\ flux in the \lya\ region, \lyalyalyalya, and its cross-correlation with \lya\ flux in the \lyb\ region, \lyalyalyalyb. The other two are cross-correlations between \lyaf\ flux and quasars, and include the cross-correlation of quasars with \lya\ flux in the \lya\ region: \lyalyaqso, and with \lya\ flux in the \lyb\ region: \lyalybqso. The \lya\ region is found between the \lya\ and \lyb\ peaks, and is defined by the rest-frame interval $\lambda_\mathrm{RF}\in [104, 120]\;$nm. The \lyb\ region is blueward of the \lyb\ peak, and is defined by the rest-frame interval $\lambda_\mathrm{RF}\in [92, 102]\;$nm. This information is summarized in \Cref{tab:corr}.

In this work, we will be extensively using the mean and covariance of the one hundred mock eBOSS correlation functions. We will also refer to this mean as the stacked correlation function. In order to compute these, we first collect all individual correlation subsamples (in HEALPix pixels) from each of the one hundred mocks. The weighted mean of all correlation subsamples over all one hundred mocks gives us the stacked correlation, while its covariance is given by the covariance of all the subsamples. This stacked correlation was computed for each of the four correlation types.

\begin{figure*}
    \includegraphics[width=1\textwidth,keepaspectratio]{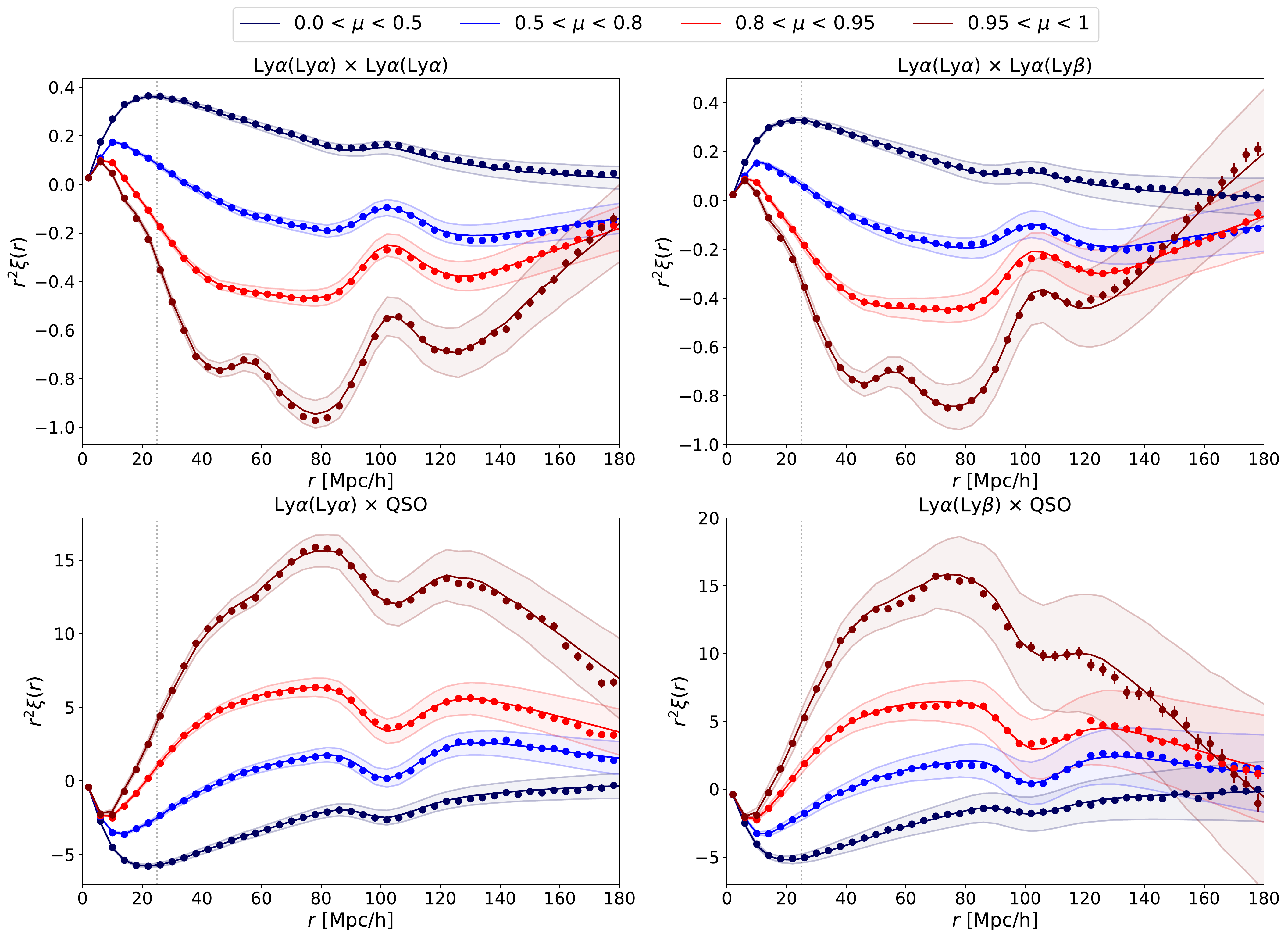}
    \caption{
    Stacked correlation functions computed from one hundred eBOSS DR16 mocks, compressed into four $\mu = r_{||}/r$ bins. The top panels show the auto-correlation of \lyaf\ transmitted flux within the \lya\ region on the left and between the \lya\ and \lyb\ regions on the right. The bottom plots show the cross-correlation between \lyaf\ transmitted flux in the \lya\ region (left) and \lyb\ region (right) and the quasar distribution. The lines represent the joint best fit model of all four correlations. The error-bars are plotted for all points, but except at large separations, they are too small to see. The shaded bands represent the uncertainties from one eBOSS realization.
    }
    \label{fig:stacked_correlations}
\end{figure*}

We show the four stacked correlations in Figure \ref{fig:stacked_correlations}, compressed into four $\mu$ bins each. Besides the BAO peak at $\sim100$\hMpc, the other features are due to metal contamination. This is why they are present in the wedges closer to the line-of-sight and absent from the ones across the line-of-sight. The most prominent of these is the blended metal peak due to SiII$(1190)$ and SiII$(1193)$, which are at comoving separations of $r_{||}=64$\hMpc\ and $r_{||}=56$\hMpc, respectively. In the auto-correlation, the SiIII$(1207)$ peak can also be seen at $r_{||}=21$\hMpc. Finally, a fourth metal peak due to SiII$(1260)$ is present at $r_{||}=111$\hMpc, however, it is not visible due to its proximity to the BAO peak.


\subsection{Modelling the correlations}
\label{subsec:model}

Our model for the \lyaf\ auto-correlation (\lyalya) and its cross-correlation with quasars (\lyaqso) closely follows that used by the eBOSS collaboration for the BAO analysis of the SDSS DR16 data \citep{duMasdesBourboux:2020}. The main difference is that we use scale parameters to model the full shape of the correlation, instead of restricting their application to scales around the BAO peak. Our model is based on an isotropic template power spectrum, $P_\mathrm{fid}(k)$, which is split into a peak (or wiggles) component, $P^\mathrm{p}_\mathrm{fid}(k)$, and a smooth (or no-wiggles) component, $P^\mathrm{s}_\mathrm{fid}(k)$, following \cite{Kirkby:2013}. As shown in \cite{Cuceu:2021}, this allows us to separate the information we obtain from the BAO peak from that obtained from the rest of the correlation. All the following steps in the analysis are done with both $P^\mathrm{p}_\mathrm{fid}(k)$ and $P^\mathrm{s}_\mathrm{fid}(k)$, in parallel. We use the \texttt{Vega} package to perform this analysis.\footnote{https://github.com/andreicuceu/vega} 

The \lya\ auto and cross power spectra are given by:
\begin{align}
    P_\mathrm{auto}(k,\mu_k,z) &= b'^2_{\mathrm{Ly}\alpha} (1 + \beta'_{\mathrm{Ly}\alpha}\mu_k^2)^2 \;G(k,\mu_k) F_\mathrm{sm}(k, \mu_k) P_\mathrm{fid}(k), \label{eq:pk_auto} \\
    P_\mathrm{cross}(k,\mu_k,z) &= b'_{\mathrm{Ly}\alpha} (1 + \beta'_{\mathrm{Ly}\alpha}\mu_k^2) \;G(k,\mu_k)  F_\mathrm{sm}(k, \mu_k) \nonumber\\
    &\times b_\mathrm{QSO} (1 + \beta_{\mathrm{QSO}}\mu_k^2) \; F_\mathrm{NL}(k,\mu_k) P_\mathrm{fid}(k), \label{eq:pk_cross}
\end{align}
where $b'_{\mathrm{Ly}\alpha}$ and $\beta'_{\mathrm{Ly}\alpha}$ are the effective \lya\ bias and RSD parameter, and $b_\mathrm{QSO}$ and $\beta_\mathrm{QSO}$ are the quasar linear bias and RSD parameter. $G(k,\mu_k) = \mathrm{sinc}(k_{||}R_{||}/2)\;\mathrm{sinc}(k_\bot R_\bot/2)$ accounts for the binning of the correlation function, with $R_{||}$ and $R_\bot$ the radial and transverse bin widths, respectively. 

The \lya\ absorption is given by a combination of contributions from the diffuse IGM and from high column density (HCD) systems. Both trace the underlying density field, but HCDs correspond to regions with significantly higher clustering. Here, we define HCDs as systems with neutral hydrogen column density above $10^{17.2} \mathrm{cm}^{-2}$, which means they include both Lyman limit systems and damped \lya\ (DLA) systems. As described in \Cref{subsec:spec_to_corr}, large HCDs are detected and masked. However, those with widths smaller than $\sim 14$\hMpc\ cannot be detected and remain in the data. These remaining HCDs lead to a broadening effect along the line of sight. \cite{Font-Ribera:2012b} showed that this can be modelled through an extra $k_{||}$ dependent term in the effective bias and RSD parameters. This is given by:
\begin{align}
    b'_{\mathrm{Ly}\alpha} &= b_{\mathrm{Ly}\alpha} + b_\mathrm{HCD} F_\mathrm{HCD}(k_{||}), \label{eq:hcd_bias} \\
    b'_{\mathrm{Ly}\alpha}\beta'_{\mathrm{Ly}\alpha} &= b_{\mathrm{Ly}\alpha}\beta_{\mathrm{Ly}\alpha} + b_\mathrm{HCD} \beta_\mathrm{HCD} F_\mathrm{HCD}(k_{||}), \label{eq:hcd_beta}
\end{align}
where $b_{\mathrm{Ly}\alpha}$ and $\beta_{\mathrm{Ly}\alpha}$ are the linear bias and RSD parameters associated with the IGM and $b_\mathrm{HCD}$ and $\beta_{\mathrm{HCD}}$ are the linear bias and RSD parameters associated with HCDs. Following \cite{Rogers:2018} and \cite{deSainteAgathe:2019}, we model $F_\mathrm{HCD}(k_{||})$ as an exponential: $F_\mathrm{HCD} = \exp(-L_\mathrm{HCD}k_{||})$. The parameter $L_\mathrm{HCD}$ can be interpreted as the typical length scale of unmasked HCDs \citep{deSainteAgathe:2019}, and is fixed to $L_\mathrm{HCD}=10$\hMpc. We test the impact of this choice in \Cref{subsec:model_tests}.

The $F_\mathrm{NL}$ term models the quasar non-linear velocities and statistical quasar redshift error. Following \cite{Percival:2009}, we test two different models for this term, one that introduces a Lorentzian damping, and one with Gaussian damping. They are given by:
\begin{align}
    F^2_\mathrm{NL,Lorentz} &= [1 + (k_{||}\sigma_v)^2]^{-1}, \\
    F^2_\mathrm{NL,Gauss} &= \exp \bigg[-\frac{1}{2}(k_{||}\sigma_v)^2 \bigg],
\end{align}
where $\sigma_v$ is a free parameter. Given that the synthetic data is created with Gaussian redshift errors, the Gaussian damping model is more appropriate for modelling the mocks. However, when modelling the real data, the Lorentz damping model is used \citep{duMasdesBourboux:2017,duMasdesBourboux:2020}. This is because the quasar redshift errors generally have long tails that are not well modelled by a simple Gaussian \citep[e.g.][]{Lyke:2020}. Therefore, for the purposes of this analysis, we will test both models.

We also introduce Gaussian anisotropic smoothing, $F_\mathrm{sm}(k, \mu_k)$, to account for the low-resolution of the \colore\ grid, following \cite{Farr:2020}. This smoothing model has two free parameters, $\sigma_{||}$ and $\sigma_\bot$, which describe the smoothing along and across the line-of-sight, respectively. They are expected to have values of $\sim2.4$\hMpc, corresponding to the resolution of the \colore\ grid.

In order to turn model power spectra into correlation functions, we follow the process described in \cite{Kirkby:2013}. This involves a multipole decomposition, followed by a Hankel transform\footnote{Performed using the FFTLog algorithm \citep{Hamilton:2000}. We use the version implemented in the \texttt{mcfit} package (https://github.com/eelregit/mcfit).} to turn each power spectrum multipole into the corresponding correlation function multipole, and finally the reconstruction of the two-dimensional correlation function from the individual multipoles. Following past \lya\ BAO analyses, we use multipole values up to $\ell=6$. The main components of the theoretical model are the \lya\ transmitted flux correlations: $\xi_{\mathrm{Ly}\alpha\times\mathrm{Ly}\alpha}$ for the auto-correlation, and $\xi_{\mathrm{Ly}\alpha\times \mathrm{QSO}}$ for the cross-correlation with quasars. They are computed from the two power spectra in \Cref{eq:pk_auto,eq:pk_cross}, respectively.

We also model the contamination due to metal absorption, using the same models as for the \lya\ auto and cross-correlation. We compute models for the correlations between each metal line and \lya\ ($\xi_{\mathrm{Ly}\alpha\times m}$), and among all metal pairs ($\xi_{m_1 \times m_2}$) for the auto-correlation, and between each metal line and quasars ($\xi_{\mathrm{QSO} \times m}$) for the cross-correlation. This is described in more detail in \Cref{app:model}. Each metal line has its own bias and RSD parameter ($b_m, \beta_m$). The same \lya\ and QSO $(b,\beta)$ parameters above are also used for the cross-correlations between metals and \lya\ and quasars. In this case, we neglect the HCD effects for the \lya\ parameters. Following past \lya\ BAO analyses, we fix the metal $\beta$ parameters to $0.5$ \citep{Bautista:2017,duMasdesBourboux:2020}.

The full model correlations are then given by:
\begin{align}
    \xi^t_\mathrm{auto} &= \xi_{\mathrm{Ly}\alpha\times\mathrm{Ly}\alpha} + \sum_m \xi_{\mathrm{Ly}\alpha\times m} + \sum_{m_1,m_2} \xi_{m_1 \times m_2}, \label{eq:xi_ta}\\
    \xi^t_\mathrm{cross} &= \xi_{\mathrm{Ly}\alpha\times \mathrm{QSO}} + \sum_m \xi_{\mathrm{QSO}\times m},  \label{eq:xi_tc}
\end{align}
where the sums are performed over the four metal lines introduced in \Cref{subsec:spec_to_corr}. The distortion due to quasar continuum errors is modelled using distortion matrices that multiply the results of \Cref{eq:xi_ta,eq:xi_tc}. See \Cref{app:model} and \cite{duMasdesBourboux:2020} for more details.

As described at the start of this section, we perform all these steps using both isotropic power spectrum components, $P^\mathrm{p}_\mathrm{fid}(k)$ and $P^\mathrm{s}_\mathrm{fid}(k)$, in parallel. Therefore, we compute the projected correlation function for both the peak component, $\hat{\xi}_\mathrm{p}$, and the smooth component, $\hat{\xi}_\mathrm{s}$. The two model correlations are then combined into the final theoretical model, allowing for their comoving coordinates to vary:
\begin{equation}
    \xi(r_{||},r_\bot) = \hat\xi_\mathrm{s}(q_{||}^\mathrm{s} r_{||}, q_\bot^\mathrm{s} r_\bot) + \hat\xi_\mathrm{p}(q_{||}^\mathrm{p} r_{||}, q_\bot^\mathrm{p} r_\bot),
\end{equation}
where $r_{||}$ and $r_\bot$ are comoving separations along and across the line-of-sight, respectively. These separations are rescaled using the scale parameters ($q_{||}^\mathrm{p}$, $q_\bot^\mathrm{p}$) for the peak, and ($q_{||}^\mathrm{s}$, $q_\bot^\mathrm{s}$) for the smooth component. The two peak scale parameters are equivalent to the ($\alpha_{||}$, $\alpha_\bot$) parameters used in BAO analyses.

In the case of the cross-correlation, we also allow for a systematic shift in the $r_{||}$ coordinate of the cross-correlation through a nuisance parameter, $\Delta r_{||}$. This means the coordinate transform is given by $r_{||} \xrightarrow{} q_{||}(r_{||} + \Delta r_{||})$. We introduce this parameter in order to find if it has any impact on the scale parameters of interest, which could lead to a potential bias when performing the analysis on real data, as it was found to be non-zero in past \lya\ BAO analyses \citep[e.g.][]{Blomqvist:2019,duMasdesBourboux:2020}.

\subsection{Modelling the Alcock-Paczy\'nski effect}

\begin{figure}
    \centering
    \includegraphics[width=0.48\textwidth,keepaspectratio]{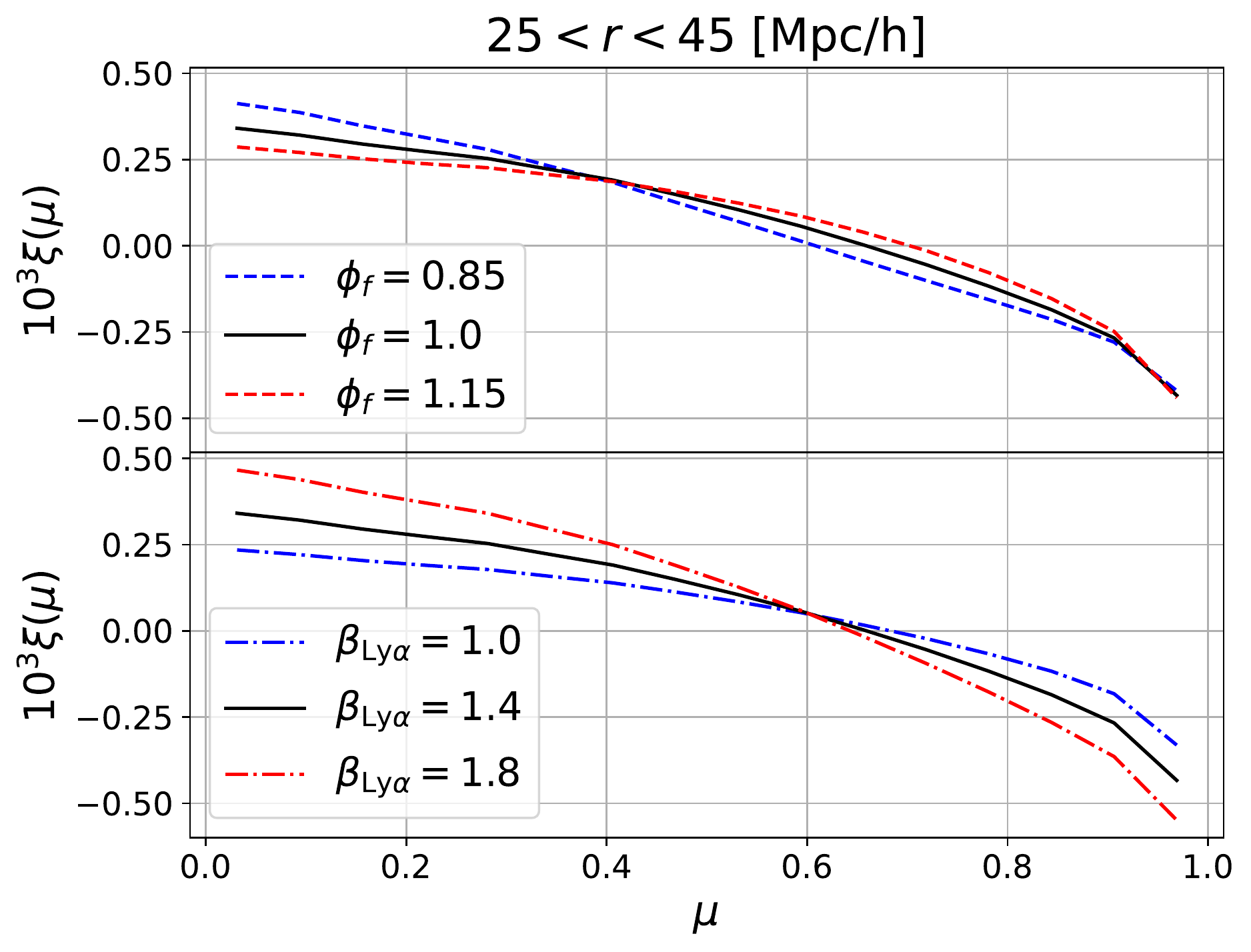}
    \caption{Models of the \lyaf\ auto-correlation function in a shell in isotropic separation $r$, shown as a function of the cosine of the line-of-sight angle, $\mu$. The top panel shows the effect of changing the value of $\phi$, the parameter measuring the Alcock-Paczy\'nski effect, while the bottom plot shows the effect of changing the value of the \lya\ RSD parameter, $\beta_{\mathrm{Ly}\alpha}$. This shows that the two effects (AP and RSD) lead to different changes in the correlation as a function of $\mu$, which explains why they are not degenerate.}
    \label{fig:phi_vs_beta}
\end{figure}

In order to isolate the Alcock-Paczy\'nski effect, we transform the scale parameter system, following \cite{Cuceu:2021}:
\begin{equation}
    \phi(z) \equiv \frac{q_\bot(z)}{q_{||}(z)} \;\;\text{and}\;\; \alpha(z) \equiv \sqrt{q_\bot(z)q_{||}(z)},
\end{equation}
where $\phi$ is the anisotropic parameter, and $\alpha$ is the isotropic scale parameter. \cite{Cuceu:2021} showed that a measurement of $\phi$ is equivalent to a measurement of 
the AP parameter $D_M(z)H(z)$. As we have two sets of ($q_{||}$,$q_\bot$) parameters, we also have two sets of ($\phi$, $\alpha$) parameters, for the peak and the smooth component, respectively. The two peak scale parameters have already been tested and measured in BAO analyses. Therefore, our goal here is to use the simulated data sets described above to study measurements of the anisotropic parameter of the smooth component, $\phi_\mathrm{s}$.

The two main types of analyses we will perform are described in \Cref{tab:analysis_types}. In the first one, which we will refer to as the "split AP measurement", we fit two distinct $\phi$ parameters for the peak and smooth components separately. In the second one, called "full AP measurement", we fit one $\phi$ parameter to the full shape of the correlations. When performing this analysis on real data, we want to use the full AP measurement, because it coherently extracts the desired information. The split AP measurement can be used as a consistency check of the method or data. However, it is also useful in the current work, because \phis\ quantifies the gain in information over a standard BAO-only analysis.

In the top panel of \Cref{fig:phi_vs_beta}, we show the impact of changing $\phi$ on the correlation function compressed in a shell in isotropic separation and plotted as a function of $\mu$. We choose the smallest separation shell, $25 < r < 45$ \hMpc, in order to best illustrate the anisotropy pattern we aim to recover when fitting the AP effect. The bottom panel of \Cref{fig:phi_vs_beta} shows the impact of changing the \lyaf\ RSD parameter $\beta_{\mathrm{Ly}\alpha}$, which also affects the anisotropy of the correlation. This figure shows that AP and RSD produce different anisotropy patterns as a function of $\mu$, which means a 2D measurement of the correlation (i.e., in $r_{||},r_\bot$ or $r,\mu$) can be used to disentangle the two effects.
 
The isotropic scale parameter from the smooth component, $\alpha_\mathrm{s}$, also contains some cosmological information due to the scale of matter-radiation equality. However, as described in \cite{Cuceu:2021}, it is not clear how to extract this information given that the effect is very similar to that produced by the \lya\ bias, and likely to be heavily impacted by contaminants. Furthermore, \cite{Cuceu:2021} found that the constraint on the isotropic scale from the peak, $\alpha_\mathrm{p}$, is much tighter than the one from the smooth component. This is in contrast to $\phi$, where the constraint from the broadband is the one that dominates (see Figure 2 of \citealt{Cuceu:2021}). Finally, \cite{Gerardi:2023} recently performed simplified forecasts of a direct cosmological analysis of the \lyaf\ correlation functions and found that the information extracted can be traced back to the Alcock-Paczy\'nski effect and redshift space distortions. Therefore, for the purposes of this analysis, we will treat $\alpha_\mathrm{s}$ as a nuisance parameter, and marginalize over it.

\begin{table*}
    \centering
    \begin{tabular}{|c|c|c|c|}
        \hline
        Analysis type & Parameters & Description & Cosmological Information \\
        \hline\hline
        \multirow{4}{7em}{Split AP measurement} & \phip & Peak anisotropy & AP from BAO \\
        & \phis & Broadband anisotropy & AP from broadband \\
        & \alphap & Peak isotropic scale & BAO scale \\
        & \alphas & Broadband isotropic scale & Marginalized out \\
        \hline\hline
        \multirow{3}{7em}{Full AP measurement} & \phif & Full-shape anisotropy & AP from full correlation \\
        & \alphap & Peak isotropic scale & BAO scale \\
        & \alphas & Broadband isotropic scale & Marginalized out \\
        \hline
    \end{tabular}
    \caption{The two types of analysis we perform in this work, along with the scale parameters we measure in each of them and their description.}
    \label{tab:analysis_types}
\end{table*}

\begin{table}
    \centering
    \begin{tabular}{ | c | m{14em} | c | }
        \hline
         Parameters & Description & Prior \\
         \hline\hline
         $\phi$, $\alpha$ & Scale parameters (see \Cref{tab:analysis_types}) & $U(0.01, 2.0)$ \\ [0.5ex]
         \hline
         $b_{\mathrm{Ly}\alpha}$ & \lya\ linear bias & $U(-2.0, 0.0)$ \\ [0.5ex]
         \hline
         $b_{\mathrm{QSO}}$ & QSO linear bias & $U(0.0, 10.0)$ \\ [0.5ex]
         \hline
         $\beta_{\mathrm{Ly}\alpha}$, $\beta_{\mathrm{QSO}}$ & RSD parameters of \lya\ and QSOs & $U(0.0, 5.0)$ \\ [0.5ex]
         \hline
         $b_{\mathrm{HCD}}$ & HCD linear bias & $U(-0.2, 0.0)$ \\ [0.5ex]
         \hline
         $\beta_{\mathrm{HCD}}$ & RSD parameter for HCDs & $\mathcal{N}(0.5, 0.2^2)$ \\ [0.5ex]
         \hline
         $\sigma_v$ [\hMpc] & Smoothing for redshift errors and QSO non-linear velocities & $U(0.0, 15.0)$ \\
         \hline
         $\Delta r_{||}$ [\hMpc] & Shift due to QSO redshift errors & $U(-3.0, 3.0)$ \\ [0.5ex]
         \hline
         $b_{\mathrm{SiII}(1190)}$ & Linear bias of metal absorber & $U(-0.02, 0.0)$ \\ [0.5ex]
         \hline
         $b_{\mathrm{SiII}(1193)}$ & Linear bias of metal absorber & $U(-0.02, 0.0)$ \\ [0.5ex]
         \hline
         $b_{\mathrm{SiIII}(1207)}$ & Linear bias of metal absorber & $U(-0.02, 0.0)$ \\ [0.5ex]
         \hline
         $b_{\mathrm{SiII}(1260)}$ & Linear bias of metal absorber & $U(-0.02, 0.0)$ \\ [0.5ex]   
         \hline
         $\sigma_{||}$ [\hMpc] & Smoothing along the line-of-sight & $U(0.0, 10.0)$ \\ [0.5ex]  
         \hline
         $\sigma_\bot$ [\hMpc] & Smoothing across the line-of-sight & $U(0.0, 10.0)$ \\ [0.5ex]  
         \hline
    \end{tabular}
    \caption{List of the free parameters in our model, along with their description and priors. $U(\text{min,max})$ represents a flat prior within that interval, while $\mathcal{N}(\mu, \sigma^2)$ represents a Gaussian prior.}
    \label{tab:priors}
\end{table}

In \Cref{tab:priors}, we give a description of all the free parameters in our standard analysis, along with their priors. We follow past BAO analyses, and assign a Gaussian prior to the RSD parameter of HCDs, based on measurements of DLA clustering \citep{Perez-Rafols:2018}.

We plot the joint best fit model using the split AP analysis configuration in Figure \ref{fig:stacked_correlations}. We fit the correlation functions between $r_\mathrm{min}=25$\hMpc\ and $r_\mathrm{max}=180$\hMpc\, following the analysis in Section \ref{subsec:scale_cuts}.\footnote{Note that we still fit for the bias parameter of the SiIII$(1207)$ line, even though that metal peak is outside our fitted range. This is because its contamination is spread all along the line-of-sight by the distortion matrix, and therefore we have to marginalize over it.} This model works surprisingly well given the extreme circumstances of this test. While the fit statistics clearly indicate the model is not good enough for fitting the stack of 100 eBOSS mocks (the fit probability is $\sim10^{-11}$), the contrast with the eBOSS uncertainties, shown as shaded bands, visually indicates that our model could be good enough for the statistical precision achievable with current data sets. We will show this using the stack of correlations in \Cref{subsec:stack_results} and by analysing individual eBOSS mocks in \Cref{subsec:mock_pop}.


\section{Results}
\label{sec:results}

A large part of the validation of our analysis will focus on studying full-shape fits using a stack of the \lyaf\ correlation functions from the eBOSS mocks. Fitting the stack of the mock correlations is much less computationally expensive compared to separately fitting all one hundred mock correlations. Therefore, this method is advantageous when we want to test the performance of our model or study different versions of the analysis. This is because it allows us to study the trends necessary without as much of an emphasis on the effect of noise. Our main goal in this section is to test if the current modelling approach for the \lyaf\ correlation functions is appropriate for full-shape analyses of the eBOSS data. In order to achieve this, we want to gain an understanding of the relation between the statistical precision we expect using the eBOSS data and the systematic bias. We also want to investigate modelling choices such as using one versus two anisotropic parameters, the relative contributions of the auto and cross-correlations, and how our modelling of the contaminants affects our measurements.


\subsection{Analysis of stacked correlations}
\label{subsec:stack_results}

\begin{figure*}
    \centering
    \includegraphics[width=0.7\textwidth,keepaspectratio]{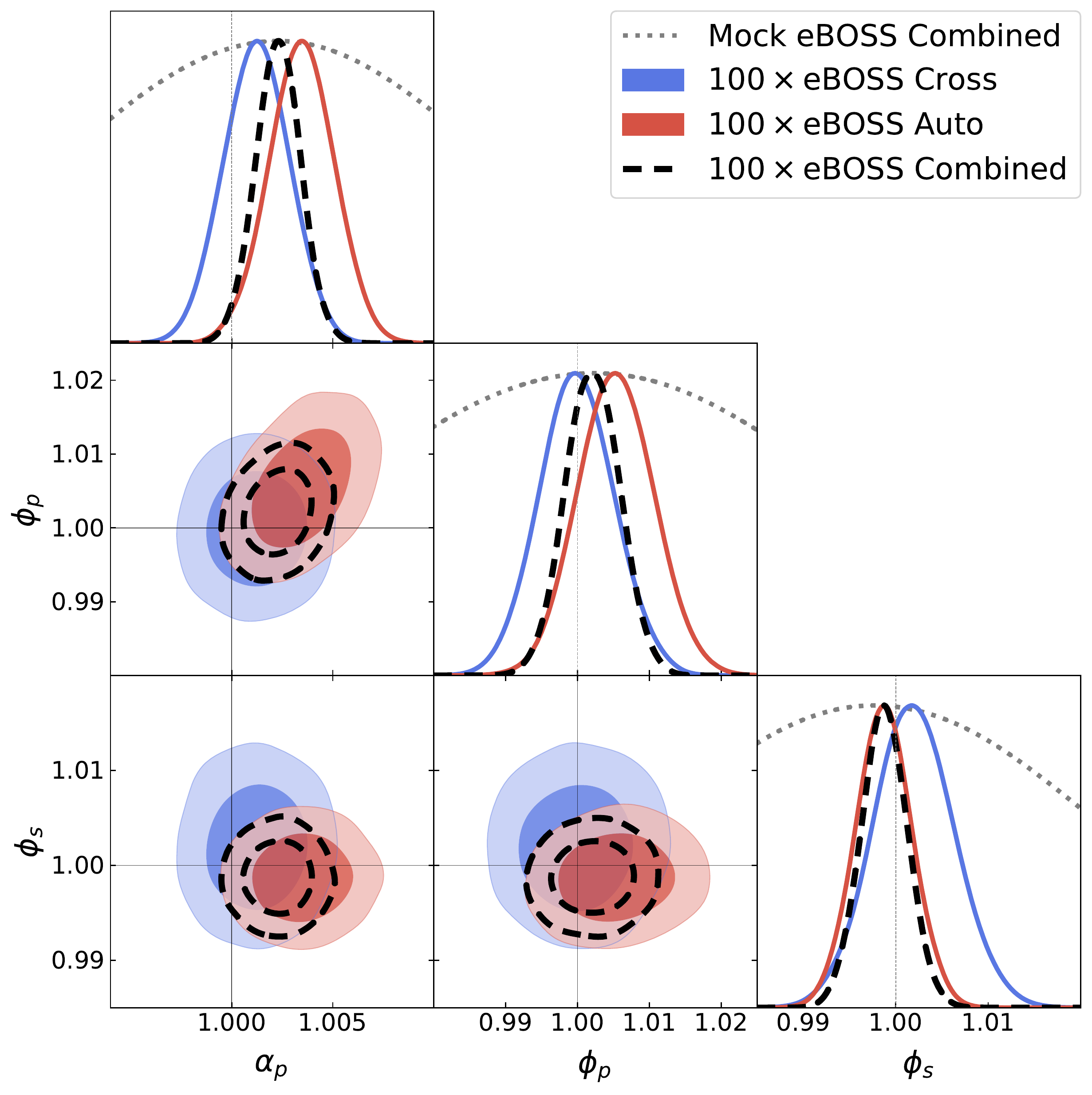}
    \caption{
    Contour plots of the constraints on scale parameters from the stack of 100 eBOSS mock correlations. The result from the two auto-correlations of \lya\ flux, \lyalyalyalya\ and \lyalyalyalyb, is denoted as "Auto", and the one from the two cross-correlations with quasars, \lyalyaqso\ and \lyalybqso, is denoted as "Cross". All scale parameters are expected to be unity, given that the same cosmology was used to construct the mocks and to analyse the data. We also show the expected eBOSS constraint for comparison, but note that the $1\sigma$ 2D contours are all larger than the bounds of the plots here. We find that we are able to recover an unbiased measurement of the AP parameter from the full shape of the \lyaf\ correlations.
    }
    \label{fig:auto_vs_cross}
\end{figure*}

As we have four different correlations, we categorise them into the two auto-correlations of \lya\ flux, \lyalyalyalya\ and \lyalyalyalyb, and the two cross-correlations with quasars, \lyalyaqso\ and \lyalybqso. This is because the modelling of the two auto-correlations is identical (and similarly for the two cross-correlations) as we treat the \lya\ flux in the \lyb\ region the same as the \lya\ flux in the \lya\ region. In the case of the two cross-correlations, the full system of parameters is degenerate. This degeneracy can only be broken when performing a joint analysis of the auto and cross-correlations, because the auto-correlation helps constrain the \lya\ bias and RSD parameters. Therefore, we fix the bias and RSD parameter of quasars ($b_\mathrm{QSO},\beta_\mathrm{QSO}$) to a fiducial value when fitting the cross-correlation alone.\footnote{The fiducial value is based on the joint constraint of all four correlations.} Note that due to correlations between the AP effect and RSD, this may lead the uncertainty in $\phi$ to be underestimated. However, we will not use these $\phi$ constraints from the cross-correlation to make any decision or recommendation in this article. We only compute and show them in order to check consistency and for completeness. Also note that, in general, the parameter constraints from the stack of the correlations will not be the same as the mean of the constraints from individual mocks. This is due to the non-linear dependence of the likelihood on the parameters. However, the stacked correlations are useful for testing the performance of our model.

We assume there is no cross-covariance between any of the four correlations, as \cite{duMasdesBourboux:2020} showed that these are negligible. We use a Gaussian likelihood, and compute posterior distributions using the \texttt{Vega} package for modelling and the \texttt{PolyChord}\footnote{https://github.com/PolyChord/PolyChordLite} package for sampling \citep{Handley:2015a,Handley:2015b}. When running \texttt{PolyChord}, we initialize a number of live points equal to 25 times the number of parameters, and set the length of the slice sampling chains to be of order of the number of parameters.

In Figure \ref{fig:auto_vs_cross}, we show the results for $\phi_s$, $\phi_p$, and $\alpha_p$ from the joint fit of the two auto-correlations alone, the two-cross correlations alone, and the combination of all four correlations, using the split AP analysis. These fits use a minimum separation $r>r_\mathrm{min}=25$\hMpc\ following the analysis in \Cref{subsec:scale_cuts}. We find good agreement between the scale parameters measured from the auto- and cross-correlations.

We use the same cosmological model that was used to construct the mocks to also compute comoving coordinates and the power spectrum template.\footnote{See \Cref{app:fid_cosmo} for tests with different fiducial cosmologies.} Therefore, we expect to recover unity for all ($\phi,\alpha$) parameters. We do recover the expected $\phi$ values from both the auto and cross-correlation, using both the peak and smooth components independently. The fact that we are able to constrain \phis\ means it is not degenerate with any of the other effects that introduce their own anisotropies. For a discussion of these effects see \Cref{subsec:model_tests}, and for plots of correlations between \phis\ and important nuisance parameters see \Cref{app:par_corr}. Finally, \Cref{fig:auto_vs_cross} shows that we are indeed able to perform an unbiased analysis of the full-shape of \lyaf\ correlations for mocks with known contaminants.

For the isotropic scale of the peak, \alphap, the result from the cross is consistent with unity. However, the result from the auto-correlation is higher, leading to the combined constraint being $\sim0.22\%$ larger than unity. While this appears as a $\sim2\sigma$ systematic bias in the analysis on the stack of 100 eBOSS mocks, it is a small fraction ($\sim0.2\sigma$) of the expected eBOSS statistical uncertainty. We will return to discuss this deviation in Sections \ref{subsec:mock_pop} and \ref{sec:discussion}.

While for the BAO peak parameters, \alphap\ and \phip\, the constraints obtained from the auto and cross-correlations are very similar, this is not the case for \phis, where the measurement from the auto-correlation is $\sim30\%$ tighter than that from the cross-correlation. This is in line with the forecasts by \cite{Cuceu:2021}. We also find that adding the correlations of \lya\ flux in the \lyb\ forest shrinks the \phis\ constraint by $16\%$ for the auto-correlation and by $10\%$ for the cross-correlation. In Figure \ref{fig:auto_vs_cross} we only focus on the parameters of interest for cosmology. However, unlike BAO parameters, \phis\ does have correlations with some nuisance parameters, the most important being $b_{\mathrm{Ly}\alpha}$, $\beta_{\mathrm{Ly}\alpha}$ and $b_{\mathrm{HCD}}$. We show these in \Cref{app:par_corr}.

\begin{figure}
    \centering
    \includegraphics[width=0.48\textwidth,keepaspectratio]{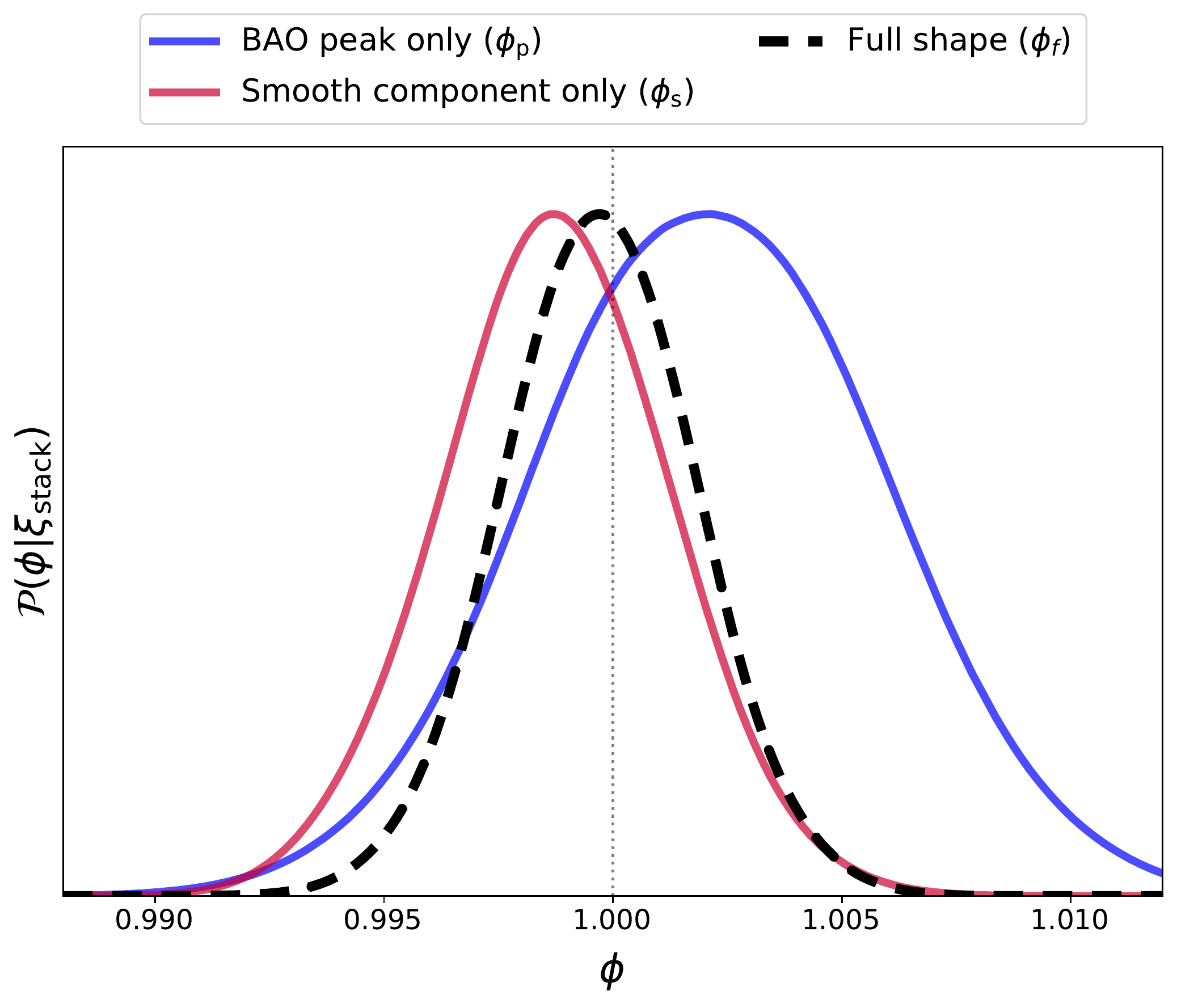}
    \caption{
    Marginalized posterior distributions of the anisotropic scale parameter $\phi$, from a joint analysis of all four \lya\ stacked correlation functions. We compare $\phi$ constraints independently measured from the peak and smooth components (denoted as \phip\ and \phis), with constraints measured from the full-shape of the correlations (denoted \phif). All measurements are consistent with the cosmology used to create and analyse the mocks, given by $\phi=1$.
    }
    \label{fig:peak_vs_smooth}
\end{figure}

In Figure \ref{fig:peak_vs_smooth}, we compare the split AP measurement with the full AP measurement, based on the joint analysis of all four stacked correlations. We show 1D marginalised posteriors of $\phi$ measurements from the peak and smooth components (\phip\ and \phis) and the measurement from the full shape (\phif). We find that the measurements from the peak and smooth components are consistent with each other, with the smooth component providing a constraint that is $\sim32\%$ tighter than the constraint from the peak component. Note that we do not expect the \phip\ and \phis\ measurements to be the same, even though they both measure the same effect. This is because they use different parts of the data, resulting in two mostly independent measurements. This is confirmed by the lack of correlation between the two parameters, as shown by their 2D posteriors in \Cref{fig:auto_vs_cross}.

The measurement of \phif\ is remarkably consistent (to within $0.1\sigma$) with the expected $\phi=1$. Overall, measuring $\phi$ from the full-shape of the correlations leads to a $\sim41\%$ improvement in the $1\sigma$ constraints compared to the BAO-only analysis. These numbers will depend on the range of scales we include in our fits, so we next turn our attention to studying the effects of the minimum scale cut we employ.


\subsection{Statistical and systematic errors}
\label{subsec:scale_cuts}

The results we have shown so far are already quite promising, with $\phi_s$ and $\phi_f$ measurements that are consistent with the true cosmology of the mocks ($\phi=1$). The next step is to quantify this statement. To achieve this, we aim to study how the statistical precision and possible systematic biases depend on the minimum scale included in our analysis. Including more data (especially at small separations) can significantly improve our constraining power \citep{Cuceu:2021}, but it may also introduce systematic biases due to imperfect modelling (of e.g. non-linear effects, HCDs, etc.). Therefore, our goal is to find a balance between these two. The key variable in this analysis is the minimum separation scale included in the fit, $r_\mathrm{min}$.

In order to quantify the systematic bias of our $\phi$ measurements, we compute the difference between the best fit value of $\phi$ from the stacked correlations and the true value in the mocks, i.e., $\Delta \phi = \phi_\mathrm{best} - 1$. While we do have the size of the statistical constraints from the stack of the correlations, we want to compare $\Delta \phi$ with the expected constraints from one eBOSS realization. Therefore, we rescale the statistical uncertainty in the fit of the stack by a factor of 10 (given the 100 mocks used to compute the stack) to obtain the expected eBOSS constraint, $\sigma_{\phi}$.

\begin{figure}
    \includegraphics[width=0.48\textwidth,keepaspectratio]{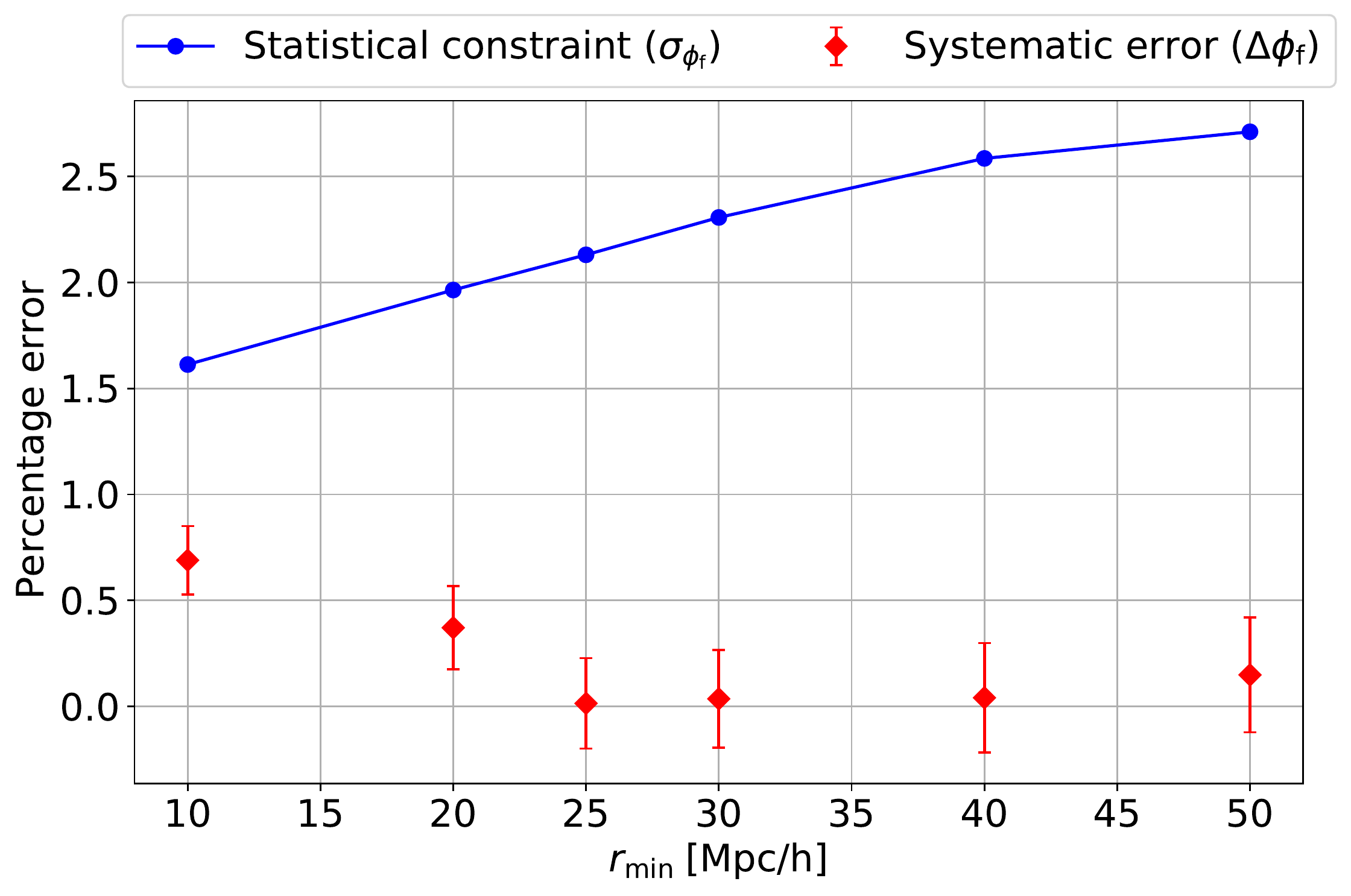}
    \caption{
    Comparison of the expected statistical constraint on $\phi$ from eBOSS ($\sigma_\phi$) and the systematic bias ($\Delta\phi$) as a function of minimum separation fitted, $r_\mathrm{min}$. The $\phi$ parameter here is fitted from the full shapes of all four \lya\ correlation functions (\phif). The systematic bias is obtained from the difference between the best fit value on the stack of 100 mocks and the true value in the mocks. The statistical constraint is obtained by rescaling the constraint from the stack of mocks to one eBOSS realization. Based on this figure, we choose $r_\mathrm{min}=25$\hMpc\ in our baseline analysis.
    }
    \label{fig:r_min}
\end{figure}

We compare the statistical constraint and the systematic error on $\phi$ for different values of $r_\mathrm{min}$ in Figure \ref{fig:r_min}. We use the full AP setup and show results for \phif\ from the joint analysis of all four correlation functions. The error-bars on the systematic errors are given by the $1\sigma$ statistical constraints of each measurement, which represents the constraining power of the stack of the one hundred mocks. 

As expected, $\sigma_\phi$ grows significantly as we cut more data (larger $r_\mathrm{min}$), with values ranging from $1.6\%$ to $2.7\%$. The systematic bias starts high for $r_\mathrm{min}=10$\hMpc\ ($0.7\%$), but becomes consistent with zero by $r_\mathrm{min}=25$\hMpc. Note that we initially performed this test in $r_\mathrm{min}$ steps of $10$\hMpc, but chose to compute the $r_\mathrm{min}=25$\hMpc\ point after seeing the initial trends. We also performed the same analysis on \phis, using the split AP approach, and found similar results (systematic bias consistent with zero for $r_\mathrm{min} \geq 25$\hMpc).

These results confirm that we are able to obtain unbiased measurements of $\phi$ using the full shapes of the \lyaf\ correlation functions in analyses using $r_\mathrm{min} \geq 25$\hMpc. Therefore, throughout the rest of this work, we will be using $r_\mathrm{min} = 25$\hMpc\ as our baseline analysis. We also studied the impact of other types of scale cuts, as described in \Cref{app:scale_cuts}. Based on that, we decided to follow BAO analyses and use a maximum scale of $r_\mathrm{max}=180$\hMpc, and not impose any anisotropic cuts.


\subsection{Modelling of contaminants}
\label{subsec:model_tests}

The results we have seen so far indicate that our current modelling of the relevant contaminants in the \lyaf\ correlations is good enough to recover unbiased full-shape measurements of the AP effect with eBOSS. Before moving to the study of analyses on individual eBOSS mocks, we wish to take a closer look at the individual components of our model. In particular, we want to understand how our measurements are affected by these individual models, and how sensitive we are to different analyses choices. Of particular interest for AP are those effects that introduce their own anisotropies.

The most important effect we have to account for is that of the distortion due to continuum fitting. This effect is modelled by introducing a projection (Equation \ref{eq:projection}; \citealt{Bautista:2017}) which removes the affected modes in both our data and our models. These distortions are limited to large scales in the power spectrum (low values of $k_\parallel)$, but affect the correlation function on all scales \citep{Blomqvist:2015,Bautista:2017}.
Given the comparison in Figure \ref{fig:stacked_correlations} between the stacked correlations and the best fit model, combined with the fact that we can recover unbiased values of \phis\ and \phif\ (Figure \ref{fig:peak_vs_smooth}), we conclude that our modelling of this distortion is accurate enough for current datasets.

While large high column density (HCD) absorbers are masked in the input spectra, those with a typical width smaller than $\sim 14$\hMpc\ remain \citep{deSainteAgathe:2019,duMasdesBourboux:2020}. The current model of this effect (Equations \ref{eq:hcd_bias} and \ref{eq:hcd_beta}) follows the description introduced by \cite{Font-Ribera:2012b}, combined with an exponential shape for the $F_\mathrm{HCD}(k_{||})$ function \citep[based on the analysis by][]{Rogers:2018}. Note that this shape assumes a model for the column density distribution function of HCDs based on observations \citep{Noterdaeme:2009,Prochaska:2010,Noterdaeme:2012,Zafar:2013}, which are very limited for column densities $\log N_{\mathrm{H}\textsc{i}}<20$. This model takes into account the joint contribution of IGM and HCD absorption at the level of two-point functions (which includes the \lya\ and HCD auto-correlations and their cross-correlation). However, \cite{Font-Ribera:2012b} found that the three-point function, $\langle \delta_{\mathrm{Ly}\alpha}\delta_\mathrm{HCD}\delta_{\mathrm{Ly}\alpha}\rangle$, also has a significant contribution. Furthermore, the process of masking the larger HCDs could also bias our correlation measurements, because we are preferentially masking regions with high clustering. These two potential sources of systematic bias have been long known \citep{Slosar:2011,Font-Ribera:2012b}. However, so far they have not had a significant impact. This appears to still be the case with full-shape analyses of eBOSS, taking $r_\mathrm{min}=25$\hMpc. However, they could be contributing to the bias we observe for $r_\mathrm{min}<25$\hMpc. We have also tested the impact of marginalizing over the $L_\mathrm{HCD}$ parameter (instead of fixing it to $10$\hMpc), and found that it has no impact on \phis\ measurements.

Contamination due to metal absorption results in clear extra features in the \lya\ correlation functions along the line-of-sight. These features correspond to correlations at vanishing separation between metal absorption and either \lya\ absorption (auto) or quasars (cross). These correlations get assigned to the wrong bins in comoving coordinates because we interpret everything as \lya\ absorption. This is a simple enough process that we can replicate it in the model through the metal matrix formalism described in \Cref{app:model}. However, it means that correlations that used to be at small separations are now spread across the correlation along the line-of-sight. This could result in potential systematic biases, as imperfect modelling of the small scales can now have an impact even on large scales. The results shown above do not indicate this is a problem for eBOSS. 

We also note that unlike HCDs, metals were added to these mocks using a re-scaled version of the \lya\ optical depth. In reality, metal absorption is associated with the circum- and intra-galactic medium more than the IGM \citep{Perez-Rafols:2022}. Therefore, these mocks only provide a rough approximation for the clustering of metals. This is especially relevant in the case of metal RSD, as the model has metal RSD parameters fixed to $\beta_\mathrm{m}=0.5$, following \cite{duMasdesBourboux:2020}. Synthetic data sets with more realistic metal prescriptions are needed to fully understand their impact on AP measurements.

Quasar redshift errors can be a significant source of uncertainty and also potentially bias both BAO and AP measurements \citep{Youles:2022}. For the auto-correlation, quasar redshifts are only used to define the rest-frame wavelength range when computing the continuum. Therefore, redshift errors result in extra spectral diversity, which increases the noise in the measurement of the flux transmission field \citep{duMasdesBourboux:2020}. On the other hand, the cross-correlation is smeared along the line-of-sight, and a potential systematic offset between negative and positive $r_{||}$ is introduced. We model the first effect using either a Gaussian or a Lorentzian model with a free parameter, as described in \Cref{subsec:model}. The second effect is modelled through an extra free parameter, $\Delta r_{||}$, to account for a potential offset. Furthermore, the Gaussian or Lorentzian models are also meant to model the non-linear quasar velocities which give rise to the Finger-of-God effect. The mocks used here do contain statistical redshift errors, modelled as a Gaussian, and the Finger-of-God effect. As analyses on real data have preferred the Lorentzian model, we tested both of them. However, we found no significant shift in any of the scale parameters of interest (\alphap, \phip, \phis).

\begin{figure}
    \centering
    \includegraphics[width=0.48\textwidth,keepaspectratio]{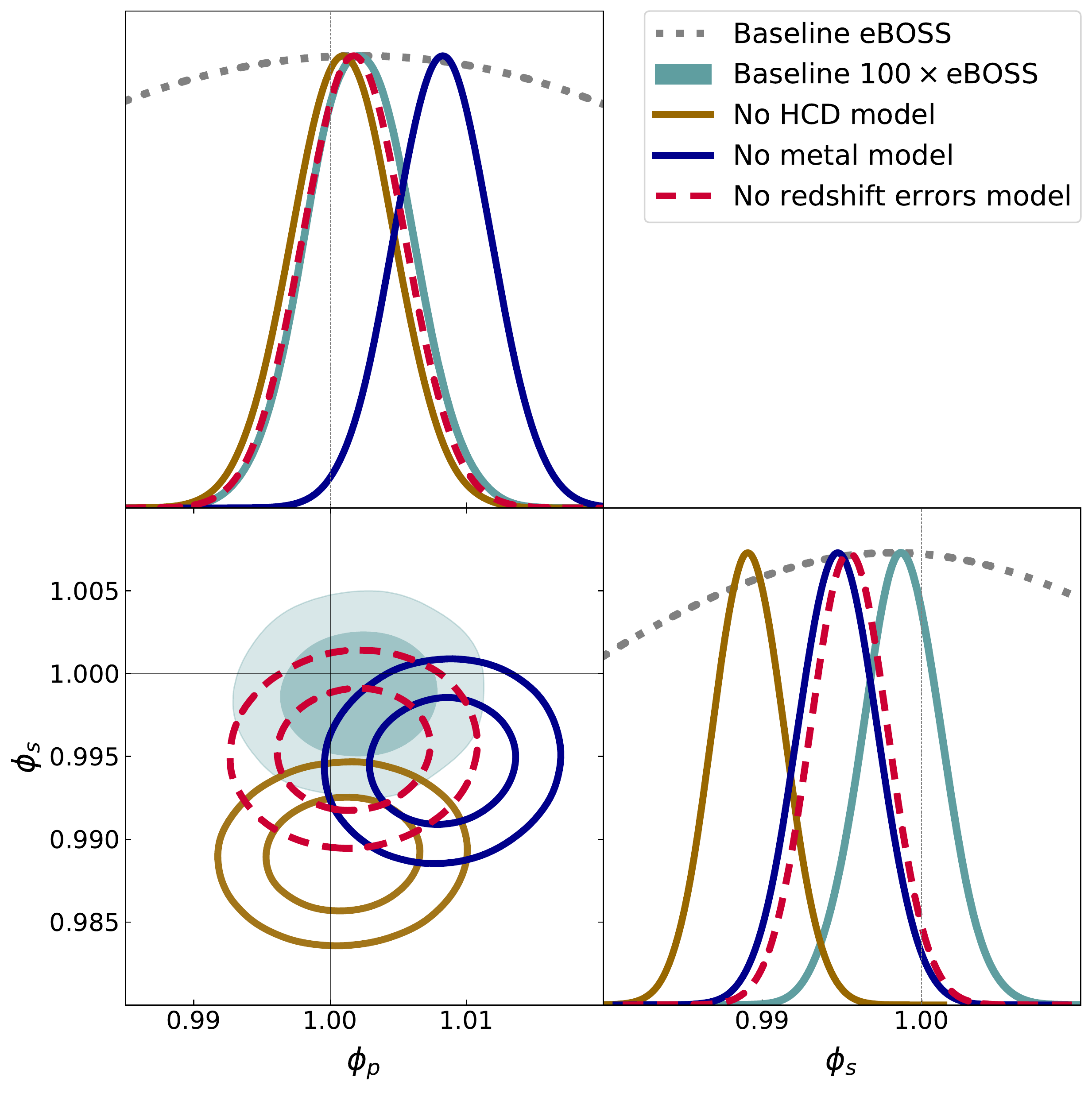}
    \caption{
    Posterior distributions of the anisotropic scale parameter, $\phi$, from the peak and smooth components, respectively. We compare the baseline result, where we model all contaminants following eBOSS \citep{duMasdesBourboux:2020}, with results where we do not model each one of the contaminants in turn. The three results here are obtained by either removing the HCD modelling, the modelling of the metals or the model for quasar non-linear velocities. In each of these cases, not modelling the relevant contaminant introduces a systematic bias in our results. For \phis, the most significant shift appears when removing the model for HCDs.
    }
    \label{fig:model_tests}
\end{figure}

In \Cref{fig:model_tests}, we compare the baseline results on $\phi$ with results where we remove the model for each of these contaminants in turn. This means that the contaminants are still present in the measured correlations, but we do not model them. These results are computed from a joint analysis of all four \lya\ stacked correlation functions. We also show the baseline result rescaled to correspond to one eBOSS realization. We study three cases: one where we remove the HCD model, one where we do not model metals, and one where we do not model quasar redshift errors. All three introduce some systematic bias in the measurement of \phis, with the largest resulting from the removal of the HCD model, leading to a $\sim 1\%$ shift. On the other hand, for \phip, only the removal of the metal modelling results in a significant bias. This figure shows the relative importance of including each of these contaminants in the model. However, note that even the largest bias (with no HCD modelling at all) represents a shift of only $\sim0.5\sigma$ of the expected statistical constraint for eBOSS.


\subsection{Analysis of individual mock realizations}
\label{subsec:mock_pop}

We now turn our attention to analysing individual eBOSS mocks. The analysis of the stacked correlations has been useful for understanding how well our model performs in the context of a full-shape analysis. We found that, using a minimum separation $r_\mathrm{min}=25$\hMpc, we are able to obtain unbiased measurements of the anisotropic scale parameter from the full shape of the correlations. However, as mentioned above, parameter constraints from the stack of mocks are not, in general, the same as mean constraints from individual mock analyses. Therefore, we now wish to test if our conclusions above are also accurate when analysing individual eBOSS mocks.


\begin{figure}
    \centering
    \includegraphics[width=0.48\textwidth,keepaspectratio]{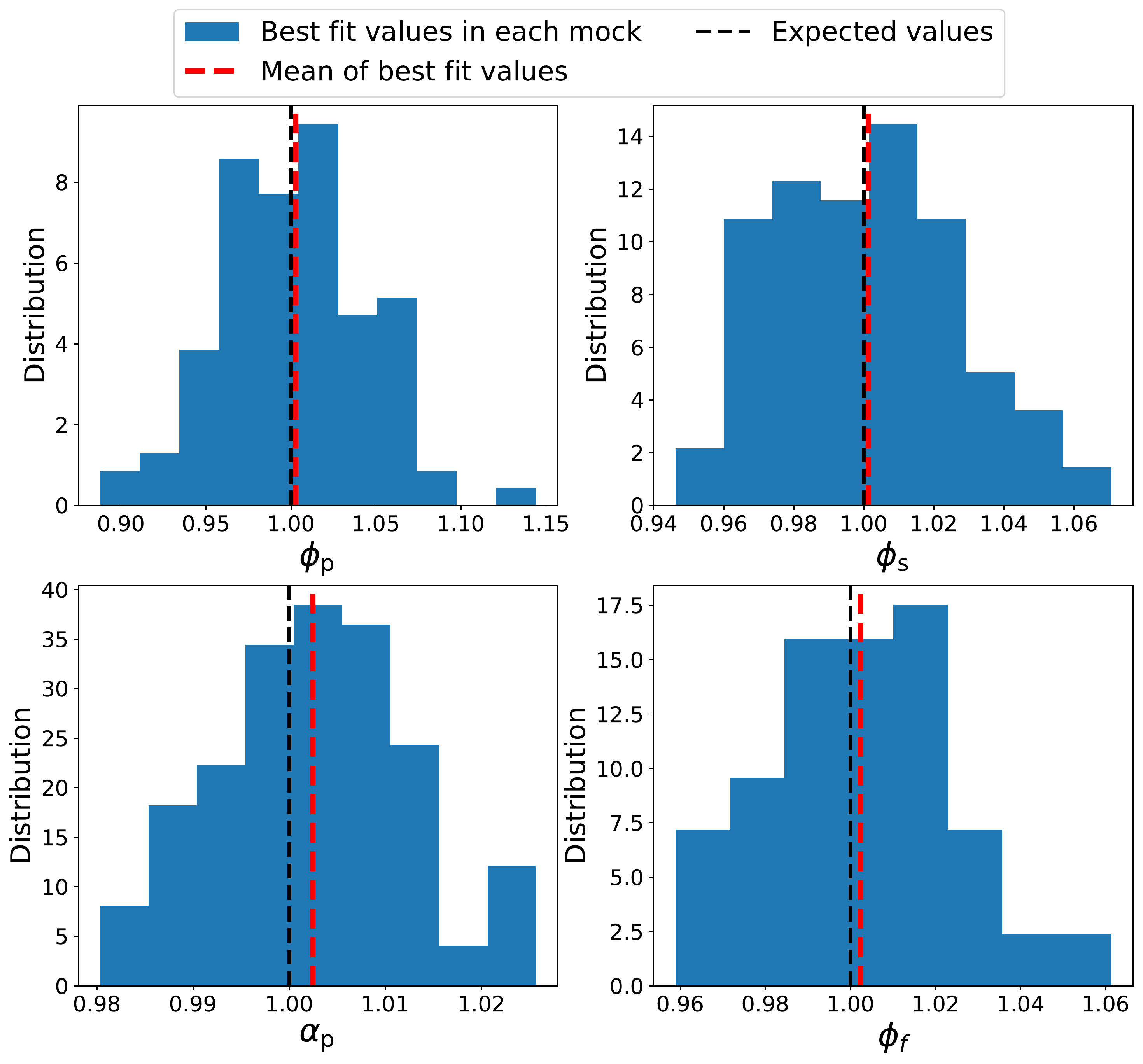}
    \caption{
    Histograms of the best fit values of scale parameters obtained by minimizing the $\chi^2$ over all parameters in each of the 100 eBOSS mocks. The red dashed lines represent the mean of the populations, while the dashed-black lines show the expected values of the scale parameters (unity). The top plots show fits of \phip\ (left) and \phis\ (right). These show that we are able to obtain unbiased measurement of $\phi$ from the full shape of the \lyaf\ correlations, which is confirmed by the bottom right plot, showing results for \phif. The bottom left plot shows results from fits of the BAO isotropic scale parameter \alphap. The mean of the population in this case is slightly higher (by $\sim0.25\sigma$) than the expected value.
    }
    \label{fig:bestfits}
\end{figure}

The results shown so far have been obtained by using a sampler to compute full posterior distributions. However, this becomes too computationally expensive for a full set of 100 mocks.\footnote{Running the sampler for a joint analysis of all four \lya\ correlations with all the contaminants requires $\sim2-3\times10^3$ CPU hours, which leads to a rough estimate of $\sim2-3\times10^5$ CPU hours to analyse all 100 mocks.} Therefore, in this section, we compute the best fit values of the parameters using a minimizer\footnote{We use the \texttt{iminuit} package \citep{Dembinski:2021} for minimization.}. We estimate the statistical uncertainty in this result using the second derivative in parameter space around the best fit point, assuming a Gaussian likelihood. This allows us to obtain rough estimates of population statistics from the set of mocks. However, note that both the best fits and the constraints obtained using this method are ultimately just an approximation, because we are not properly marginalizing over the nuisance parameters (see \citealt{Cuceu:2020}). We only use this method here because a more complex and accurate method (i.e. sampling) is not practical. We discuss this in more detail in \Cref{app:mock_constraints}.

Our modelling of individual eBOSS mock correlations closely follows the model we used for the stack of correlations, as introduced in Section \ref{subsec:model}. We run the \texttt{iminuit} minimizer on each of the one hundred mocks. This provides us with both the best fit value of each parameter and with its Gaussian uncertainty. We compute two versions of the results, using both the split AP analysis with \phip\ and \phis, and the full AP analysis with \phif.

We show histograms of the best fit values of the scale parameters in Figure \ref{fig:bestfits}. The top plots show results for \phip\ and \phis, both of which are consistent with the expected value of unity. The means of the populations are $\mu_{\phi_\mathrm{p}}=1.0028$ and $\mu_{\phi_\mathrm{s}}=1.0013$, corresponding to shifts from unity of $\sim0.06\sigma$ and $\sim0.05\sigma$, respectively.\footnote{The $\sigma$ values here are based on the expected eBOSS uncertainty given by the standard deviation of these histograms. This is also consistent with the scaled uncertainty from the stack of mocks.} These results reinforce the conclusions from \Cref{subsec:stack_results}: we are indeed able to recover unbiased measurements of the AP effect using the full shapes of the correlations.

The bottom row of Figure \ref{fig:bestfits} shows results for the isotropic BAO scale (\alphap) and the full-shape anisotropic parameter (\phif). The means of the populations in this case are $\mu_{\alpha_\mathrm{p}}=1.0024$ and $\mu_{\phi_f}=1.0024$, corresponding with shifts from unity of $\sim0.23\sigma$ and $\sim0.11\sigma$, respectively. While these shifts are more significant compared to the ones from \phip\ and \phis, they are still relatively small, in line with the results from \Cref{subsec:stack_results}.


\section{Discussion}
\label{sec:discussion}

In this work, we have used a set of one hundred eBOSS \lyaf\ mocks to study how well we could perform a full-shape analysis of the eBOSS DR16 \lya\ correlation functions. We found that we are able to recover unbiased measurements of the anisotropic scale parameter, $\phi$, from a joint analysis of the full shapes of the four \lya\ correlations when using a minimum separation of $r_\mathrm{min}=25$\hMpc. The only significant systematic biases on $\phi$ were found when either we did not model one of the major contaminants (\Cref{subsec:model_tests}), or we included scales smaller than $25$\hMpc\ (\Cref{subsec:scale_cuts}).

We did find an apparent bias of $0.22\%-0.24\%$ in the isotropic scale of the peak component, \alphap, both when analysing the stack of one hundred mocks (\Cref{subsec:stack_results}) and when analysing them individually (\Cref{subsec:mock_pop}). This corresponds to a $\sim 0.21\sigma-0.25\sigma$ bias for eBOSS.\footnote{Note that the eBOSS uncertainty used for these values are based on estimates from this work, not the real measurement.} \cite{duMasdesBourboux:2020} also found a systematic shift of similar magnitude for the BAO parameters, using the same correlation function measurements with a different parametrization. In the same publication, \cite{duMasdesBourboux:2020} measured BAO using real data, and tested the robustness of the result by adding broadband polynomials meant to marginalize over any unaccounted broadband systematic effects. They found no significant shift in the BAO parameters when performing this test. Therefore, they concluded the BAO results are robust. Such a test cannot be replicated in the context of a full-shape analysis, because the broadband polynomials would erase the information in \phis. However, our model allows the separate fit of BAO (\alphap\ and \phip) and broadband (\phis) information. Therefore, when analysing real data, we can benchmark our BAO result using the results of the eBOSS BAO analysis.

In \Cref{subsec:model_tests}, we have discussed the impact of the major contaminants that are modelled in the mock data and shown the importance of modelling them accurately. However, we have only touched on the effects that we model and observe in the mock data. While these represent the most important known contaminants for the \lyaf\ 3D correlations (which is why they are modelled in the mock data), they are not the only possible sources of systematic uncertainty. Here we wish to discuss some other possible sources of contamination, how they have been treated in the past, and also to give recommendations on which of them warrant attention when performing this analysis on real data.

The mock data used in this work is based on a Gaussian density field with quasars randomly sampled using its log-normal transformation. While this is a good enough approximation on large scales, it does not correctly reproduce the quasar clustering on small scales \citep{Farr:2020}. This makes it difficult to accurately test the model on small scales, especially for the cross-correlation. The problem is also exacerbated by the fact that both the distortion due to continuum fitting and the metal contamination take effects that are at small scales and spread them to larger scales along the line-of-sight. Incorrect modelling of the small scales can thus have an impact on all scales, so this analysis would benefit from more realistic mocks to test this effect. However, as these are not currently available, we have to leave such a study for future work.

For the analysis of the \lya\ auto-correlation from real data, an empirical model based on the work by \cite{Arinyo:2015} is used to fit small-scale non-linearities.\footnote{We do not use this model here because the small-scale forest-related non-linearities are not included in the mocks.} This model has five free parameters whose values were measured using hydro-simulations. However, no equivalent model exists for the cross-correlation. A possible approximation would be to use the same model for the cross-correlation as well, but it is not clear if the same parameter values would still work, and this would not account for quasar non-linearities (FoG). Recently, \cite{Givans:2022} used high-resolution hydrodynamic simulations to study how well current models perform on small scales in both the \lya\ power spectrum and the cross-spectrum of \lya\ with massive halos. For \lya, they found that the model by \cite{Arinyo:2015} performs very well up to wavenumbers much larger than those included here. On the other hand, for the cross-spectrum, they found that the standard Lorentzian damping combined with the \cite{Arinyo:2015} model cannot accurately fit the small-scale power. Therefore, for the cross-correlation, they recommend using scales larger than $30$\hMpc\ when fitting it with linear theory. However, they did not include the effect of redshift errors, which suppresses power on these small scales, making deviations from linear theory less important. Therefore, in an analysis of real data, it may be worth testing scale cuts for the cross-correlation of both $25$\hMpc, as proposed here, and of $30$\hMpc, as proposed by \cite{Givans:2022}.

Comparing results from the auto and cross-correlations could be a good way of testing for systematic biases due to imperfect modelling of the small scales. This is because the two are driven by different effects, with the cross-correlation affected more by redshift errors and quasars, which cluster very strongly and have large non-linear peculiar velocities, and the auto-correlation driven by the IGM absorption (gas pressure effects, thermal broadening, etc.). Furthermore, for the cross-correlation, testing both the Gaussian and Lorentzian damping models would be a good consistency check for the analysis on real data.

While preparing this manuscript, a new study by \cite{Youles:2022} found that large redshift errors can introduce spurious correlations along the line-of-sight in both the \lya\ auto and cross-correlation with quasars. Such an effect could lead to systematic biases when performing a full-shape analysis such as the one studied here. Therefore, future work is needed to study its impact on full-shape analyses, and to potentially find ways to account for it (e.g. by forward modelling the effect).

An effect that is not included in the mock data is that of BAO broadening due to non-linear growth. However, this effect is relatively well understood and modelled in BAO analyses (see \citealt{Kirkby:2013} for a detailed discussion). Therefore, as long as it is correctly modelled when fitting the real data, we do not consider it a potential issue for a full-shape analysis. The effect of quasar radiation, also known as the transverse proximity effect, is another important source of contamination for the cross-correlation, because it increases the ionization fraction in the gas surrounding the quasar, leading to a decrease in \lya\ absorption. However, this effect has been successfully modelled \citep{Font-Ribera:2013}, and this model has been included in previous BAO analyses. As this only affects the cross-correlation, comparing results from the auto- and cross-correlation should also be useful here.

The metal contamination introduced in the synthetic data used here only includes the effect of the lines that are close in wavelength to the \lya\ peak. This type of contamination is mostly given by the cross-correlation between these absorptions lines and either \lya\ absorption (for the auto-correlation) or quasars (for the cross-correlation). However, in BOSS and eBOSS BAO analyses, the contamination due to CIV absorption was also modelled. At the relevant comoving coordinates used here, this is dominated by the auto-correlation of CIV absorption, and therefore only affects the \lya\ auto-correlation. It is generally modelled in the same way as the other metal lines, as described in \Cref{subsec:model}. It has its own free bias parameter. However, this parameter has only been marginally detected, with a significance less than $3\sigma$ \citep{duMasdesBourboux:2020}. Therefore, the impact of this contamination is minimal. Nevertheless, we recommend that it be modelled and marginalized over in a full-shape analysis on real data, similarly to \lya\ BAO analyses.

Fluctuations of the ionizing UV background introduce a scale dependence to the \lya\ bias and RSD parameters \citep{Pontzen:2014,Gontcho:2014}. This scale dependence was modelled in the \lya\ BAO analyses using BOSS DR12 and eBOSS DR14 \citep{Bautista:2017,deSainteAgathe:2019,Blomqvist:2019}, following the framework introduced by \cite{Gontcho:2014}. However, the effect was not detected at a significant level (only $\sim2\sigma$), and modelling it did not have an impact on the BAO fits. This indicates that it is probably not of significant concern for a full-shape analysis, but its impact on $\phi$ measurements should be tested when performing this analysis on real data.

In conclusion, the two effects that likely require more attention are the spurious correlations due to redshift errors \cite[identified by][]{Youles:2022}, and the small scale non-linear effects in the cross-correlation. The other effects discussed here have been studied before, and modelled at some level in past \lya\ BAO analysis. They were not prioritized by the eBOSS collaboration when creating mocks because they were not found to play an important role in past BAO analyses. However, this may not be true for full-shape analyses. Therefore, an important part of the first full-shape analysis on real data will be to determine if the current models for these secondary effects have an impact on measurements of \phis. If they are found to have an impact, it will motivate the development of more accurate mocks in the future.


\section{Summary}
\label{sec:conclusion}

The Lyman-$\alpha$ (\lya) forest provides one of the best tracers of large scale structure (LSS) at high redshifts ($2<z<4$). However, 3D correlations of the forest have so far only been used to measure the baryon acoustic oscillations (BAO) scale. Recently, \cite{Cuceu:2021} showed that a measurement of the Alcock-Paczy\'nski (AP) effect from the full shapes of \lyaf\ correlation functions can lead to significant improvements in cosmological constraints compared to BAO-only measurements. In this work, we use synthetic data of the extended Baryon Oscillation Spectroscopic Survey (eBOSS) in order to test such an analysis.

We model the \lya\ auto-correlation, and its cross-correlation with quasars, using a similar approach to that used in past \lya\ BAO analyses \citep{Kirkby:2013,Bautista:2017,duMasdesBourboux:2017,duMasdesBourboux:2020}, as described in \Cref{subsec:model}. The only difference appears from the fact that we use scale parameters to fit the full shapes of the correlations. Our parametrization of the scale parameters splits them into an isotropic scale parameter, $\alpha$, and a parameter for the anisotropy, $\phi$, corresponding to a measurement of the AP effect \citep{Cuceu:2021}. As we decompose the template power spectrum into a peak component and a smooth component, we have the option of using separate scale parameters for each. We use two setups for the analysis, as described in \Cref{tab:analysis_types}. In the first, we decouple the peak and smooth components by fitting two distinct $\phi$ parameters (\phip\ and \phis), while in the second we fit only one $\phi$ parameter (\phif) for the full shapes of the correlations.

We begin our study by analysing stacked correlation functions from the one hundred mock data sets (see \Cref{fig:stacked_correlations}), in \Cref{subsec:stack_results}. We perform a joint analysis of all four correlations, as shown in Figures \ref{fig:auto_vs_cross} and \ref{fig:peak_vs_smooth}. We find that the results from the two \lya\ auto-correlations are consistent with the results from the two cross-correlations. Furthermore, we find that we are able to recover unbiased measurements of $\phi$ from the full shapes of the correlations, for both types of analysis described above.

In \Cref{subsec:scale_cuts}, we study joint analyses of all four stacked correlations using a range of scale cuts. In particular, we focus on the minimum scale used, $r_\mathrm{min}$. We run the full AP analysis using $r_\mathrm{min}$ values from $10$\hMpc\ to $50$\hMpc, as shown in \Cref{fig:r_min}. We then compare the systematic bias of \phif\ measurements, given by the difference between the best fit value and the expected value in the mocks ($\phi=1$), with an estimate of the expected statistical constraint from eBOSS. We find that we are able to obtain unbiased measurements of \phif\ for $r_\mathrm{min}\geq25$\hMpc. Therefore, we use $r_\mathrm{min}=25$\hMpc\ for the other parts of the analysis.

In \Cref{subsec:model_tests}, we discuss the main contaminants present in the \lyaf\ correlation functions. We also test a few different setups for the analysis, but find that as long as the main contaminants are modelled correctly, there is no significant bias in full-shape $\phi$ measurements. We show the relative impact of not modelling each main contaminant in \Cref{fig:model_tests}.

We study full-shape analyses on the individual correlations in \Cref{subsec:mock_pop}. We used approximate results from a minimizer instead of running a sampler on each of the one hundred mocks, due to computational constraints (see \Cref{app:mock_constraints} for a comparison between the two). We show histograms of the best fit parameter values in \Cref{fig:bestfits}. Our results here reinforce the earlier conclusions. The means of these distributions are consistent with the values expected from the mocks, $\phi=1$. This again indicates that we are able to obtain unbiased measurements of $\phi$, using the full shapes of the correlations.

Finally, in \Cref{sec:discussion} we discuss the implications of our results and give recommendations for how such an analysis should be performed on real data from eBOSS. As we have shown, the most important contaminants are well modelled and do not have a significant impact on $\phi$ measurements for $r_\mathrm{min}\geq25$\hMpc. Therefore, we focus on discussing the effects that are not modelled in the synthetic data, and what tests can be done on the real data in order to gauge their impact. Most of these are quite well understood, have already been modelled and tested in past \lya\ BAO analyses, and were not found to have a significant impact. However, these tests should also be replicated in the context of a full-shape analysis.

An eBOSS full-shape analysis, as studied here, could improve AP constraints by as much as a factor of two compared to a BAO-only analysis. We have performed this measurement in a follow-up work presented in \cite{Cuceu:2022}. While our focus here has been on the information extracted through the AP effect, our validation of full-shape measurements could also prove important to other sources of information present in the broadband of \lyaf\ correlations. These include growth of structure measurements using redshift space distortions \citep{Cuceu:2021}, and fluctuations in the ionising UV background \citep[e.g.][]{Long:2023}.

\section*{Data Availability}

Data supporting this research is available on reasonable request from the corresponding author. 


\section*{Acknowledgements}

AC and PM acknowledge support from the United States Department of Energy, Office of High Energy Physics under Award Number DE-SC-0011726. AFR acknowledges support through the program Ramon y Cajal (RYC-2018-025210) of the Spanish Ministry of Science and Innovation and from the European Union’s Horizon Europe research and innovation programme (COSMO-LYA, grant agreement 101044612). 
IFAE is partially funded by the CERCA program of the Generalitat de Catalunya. 
BJ acknowledges support by STFC Consolidated Grant ST/V000780/1.
SN acknowledges support from an STFC Ernest Rutherford Fellowship, grant reference ST/T005009/2. 
We acknowledge the use of the \texttt{GetDist} \citep{Lewis:2019} and \texttt{numpy} \citep{Harris:2020} packages. 
For the purpose of open access, the authors have applied a CC BY public copyright licence to any Author Accepted Manuscript version arising.



\bibliographystyle{mnras}
\bibliography{main} 




\appendix

\section{Modelling metals and the distortion}
\label{app:model}

As we assume all forest pixels correspond to \lya\ absorption, correlations involving metals are assigned to the wrong bin in comoving coordinates due to one (for Ly$\alpha\times m$ and QSO$\times m$) or both (for $m\times m$) tracers being assigned the wrong redshift. This is modelled through a metal matrix which transforms the model correlations computed using the correct separations, $\Tilde{r}_{||}$ and $\Tilde{r}_\bot$, to the measured model correlations:
\begin{equation}
    \xi^\mathrm{metal}_A = \sum_B M_{AB} \xi^\mathrm{metal}(\Tilde{r}_{||}(B), \Tilde{r}_\bot(B)),
\end{equation}
where $A$ and $B$ are comoving separation bins, and the metal matrix is given by:
\begin{equation}
    M_{AB} = \frac{1}{W_A} \sum_{(i,j) \in A,(i,j) \in B} w_i w_j,
\end{equation}
where $(i,j) \in A$ refers to separations computed using the assigned redshifts (assuming only \lya\ absorption), and $(i,j) \in B$ refers to separations computed using the true redshifts of each tracer. The weights, $w_i$, are given by the inverse variance of $\delta_q$ (see \citealt{duMasdesBourboux:2020}), and $W_A = \sum_{(i,j)\in A} w_i w_j$. Note that we drop the quasar specification from our notation, as the sums are understood to be over all pixels in each comoving separation bin.

In order to model the distortion due to continuum fitting errors, we need to apply the projection $\eta_{ij}^q$ to the modelled flux fluctuation $\delta^t_q$ (Equation \ref{eq:projection}). This is done by using $\eta_{ij}^q$ to construct a distortion matrix that is applied to the theoretical model correlations, $\xi_\mathrm{auto}^t$ and $\xi_\mathrm{cross}^t$ to give the projected correlation function:
\begin{equation}
    \hat\xi_A = \sum_{B} D_{AB}\xi_{B}^t,
\end{equation}
where $A$ is a bin of the data and $B$ is a bin of the model. The distortion matrices for the auto and cross-correlations are given by:
\begin{align}
    D_{AB}^\mathrm{auto} &= W_A^{-1} \sum_{ij \in A} w_i w_j \sum_{i'j' \in B} \eta_{ii'} \eta_{jj'}, \\
    D_{AB}^\mathrm{cross} &= W_A^{-1} \sum_{ij \in A} w_i w_j \sum_{i'j \in B} \eta_{ii'}.
\end{align}
See \cite{Bautista:2017} and \cite{Perez-Rafols:2018} for more details.

\section{Impact of fiducial cosmology}
\label{app:fid_cosmo}

Our method requires the use of a fiducial cosmology, both for defining the comoving coordinate grid on which the correlations are computed, and for the template power spectrum that is used to construct the theoretical model for the \lya\ correlations. This fiducial cosmology is based on cosmic microwave background anisotropy data from the Planck satellite \citep{Planck:2016}. In this appendix, we test the impact of this choice of fiducial cosmology.

We recompute the \lya\ auto and cross-correlations of one of the mocks using two different values of the matter density parameter, $\Omega_m$. The two values are $\Omega_m=0.27$ and $\Omega_m=0.36$. We then perform our standard analysis and measure \alphap\ and \phif, but we use different power spectrum templates to keep the fiducial $\Omega_m$ value consistent in each analysis. In order to compare the results, we compute the relevant combinations of distances as derived parameters. Following \cite{Cuceu:2021}, measurements of $\phi$ and \alphap\ correspond to:
\begin{align}
	\textbf{AP: } \; \phi(z) &= \frac{D_\mathrm{M}(z) H(z)}{[D_\mathrm{M}(z) H(z)]_\text{fid}}, \label{eq:phi} \\
	\textbf{BAO: } \; \alpha_\mathrm{p}(z) &= \sqrt{\frac{D_\mathrm{M}(z) D_\mathrm{H}(z) / r_\mathrm{d}^2}{[D_\mathrm{M}(z) D_\mathrm{H}(z) / r_\mathrm{d}^2]_\text{fid}}},
    \label{eq:alpha}
\end{align}
where $D_H = c/H$ and $r_d$ is the size of the sound horizon at the end of the drag epoch. Based on these equations and the relevant fiducial distances for each value of $\Omega_m$, we transform \phif\ into $D_M / D_H$, and \alphap\ into $D_M D_H / r_d^2$.

\begin{figure}
    \centering
    \includegraphics[width=0.48\textwidth,keepaspectratio]{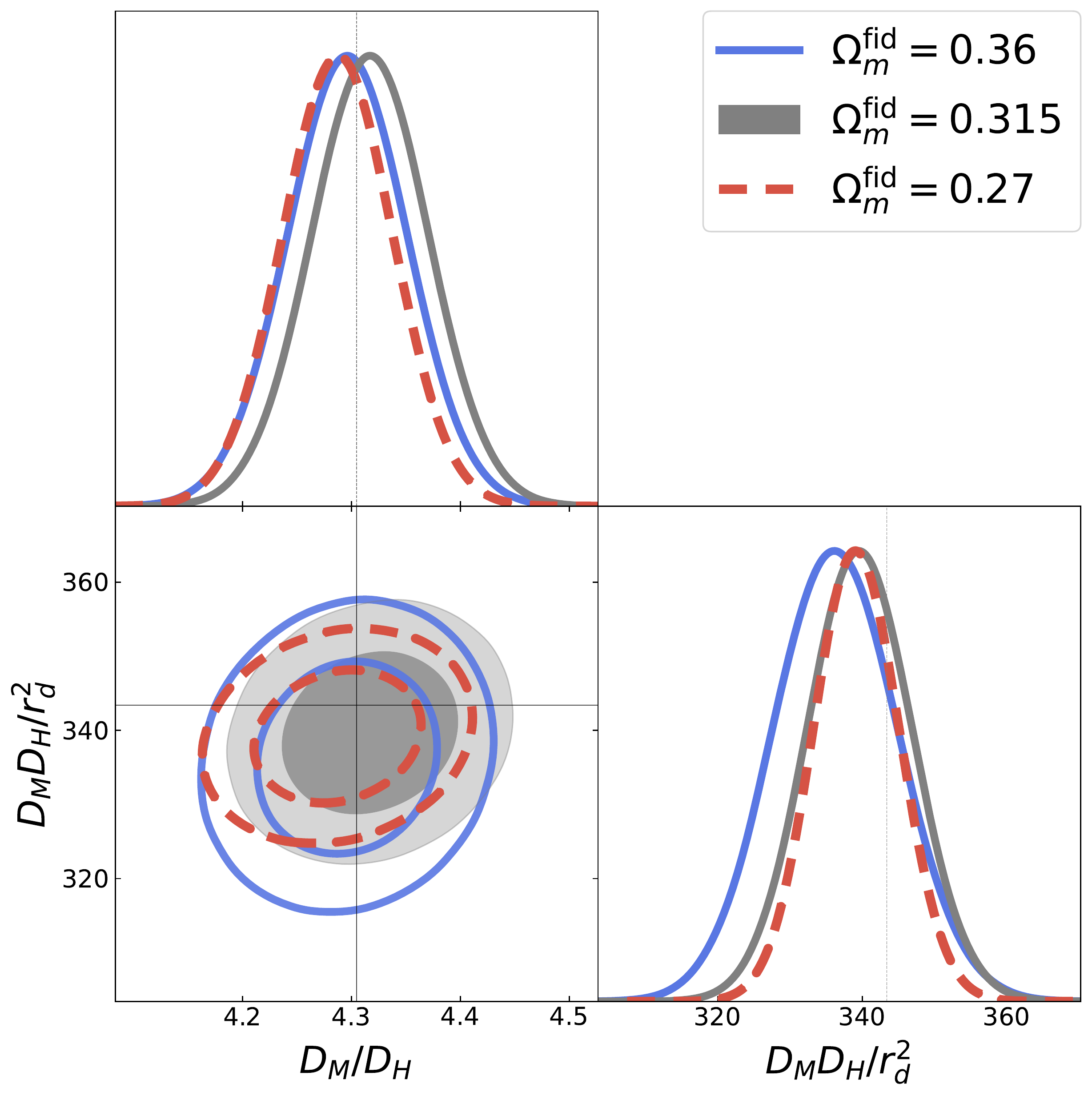}
    \caption{Posterior distributions of the combination of distances we measure for different $\Omega_m$ values used in the fiducial cosmology. The combination $D_M/D_H$ corresponds to \phif, while $D_M D_H / r_d^2$ corresponds to \alphap. This shows that we measure consistent distances even when using a very different fiducial cosmology.}
    \label{fig:omega_m}
\end{figure}

We compare the results for the large and small values of $\Omega_m$ with the one obtained from our standard analysis (using $\Omega_m=0.315$) in Figure \ref{fig:omega_m}. All three results give consistent measurements, however, they are not identical. This can be attributed to the fact that pixel pairs are distributed to different bins in the correlation when using different values of $\Omega_m$, because the comoving distances change. This will have an impact on the measured parameters, especially for BAO, where small changes in the correlation measurements (consistent with the expected noise) can produce large changes in the shape of the posterior \citep[e.g.][]{Cuceu:2020}. Furthermore, when changing the fiducial $\Omega_m$, the smallest scale that we fit will be different for the same $r_\mathrm{min}$. For example, the smaller value of $\Omega_m$ leads to smaller scales being assigned larger comoving separations. This is why the size of the constraints increases as $\Omega_m$ increases.\footnote{This also means that our choice of $r_\mathrm{min}=25$\hMpc\ is based on the Planck cosmology. Therefore, the analysis should be re-done, and a new minimum scale chosen, if we had reason to believe the real cosmology was very different from the one inferred by Planck.} Nevertheless, all three results are consistent with each other even for these extreme values of $\Omega_m$.

\section{Other scale cuts}
\label{app:scale_cuts}

\begin{figure}
\centering
    \centering
    \includegraphics[width=0.48\textwidth,keepaspectratio]{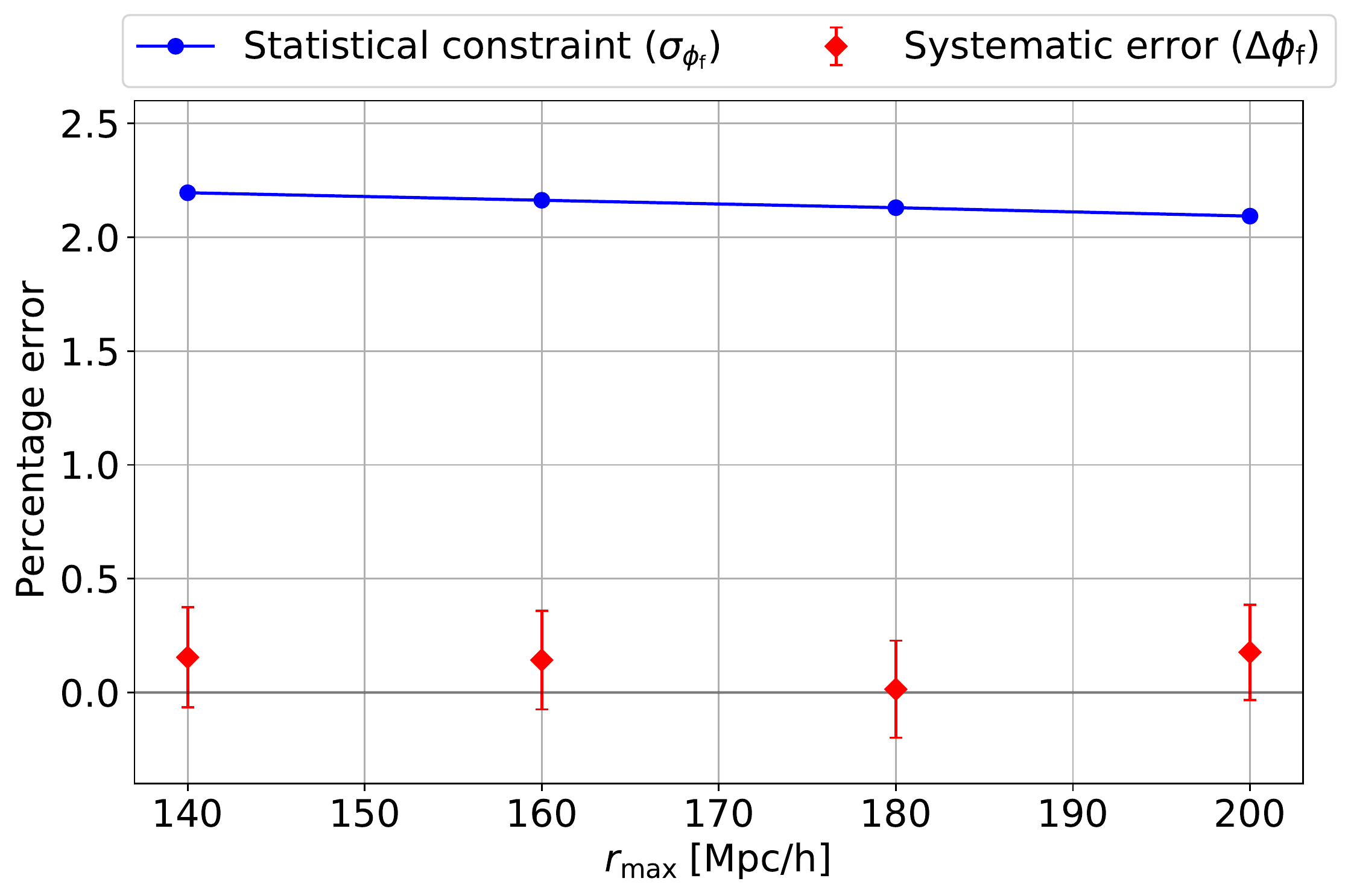}
    \caption{Comparison of the expected statistical constraint on $\phi$ from eBOSS ($\sigma_\phi$) and the systematic bias ($\Delta\phi$) as a function of maximum scale fitted, $r_\mathrm{max}$. The $\phi$ parameter here is fitted from the full shapes of all four \lya\ correlation functions (\phif). The systematic bias is obtained from the difference between the best fit value on the stack of 100 mocks and the true value in the mocks. The statistical constraint is obtained by rescaling the constraint from the stack of mocks to one eBOSS realization.}
    \label{fig:rmax}
\end{figure}

\begin{figure}
    \centering
    \includegraphics[width=0.48\textwidth,keepaspectratio]{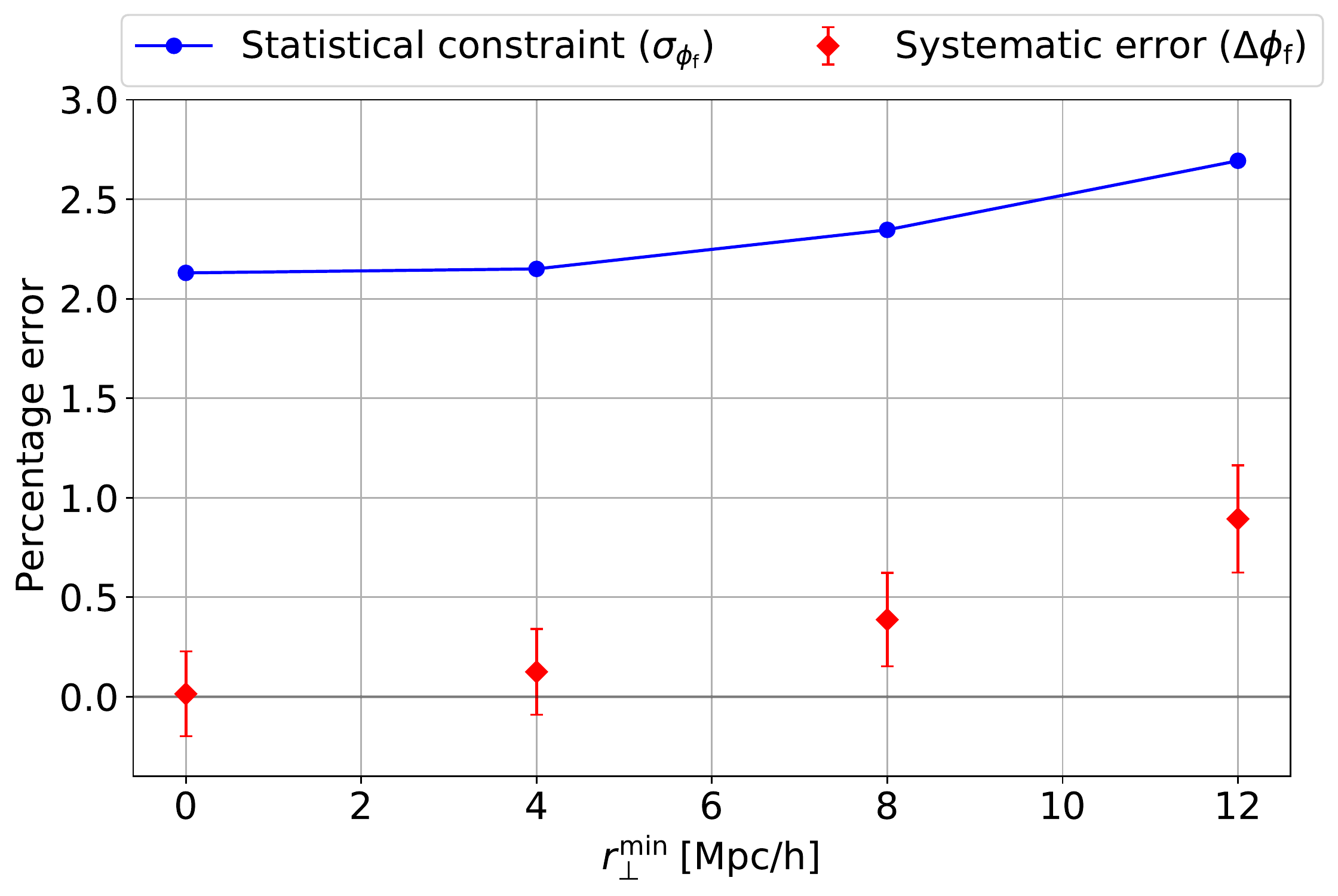}
    \caption{Same as Figure \ref{fig:rmax}, but as a function of the minimum scale across the line-of-sight, $r_\bot^\mathrm{min}$. We are progressively cutting the columns along the line-of-sight, where the impact of the contaminants is highest. The systematic error on $\phi$ is increasing as we cut more data, which indicates that we are not able to accurately constrain the nuisance parameters that account for the contaminants when ignoring the information along the line-of-sight.}
    \label{fig:rtmin}
\end{figure}

In \Cref{subsec:scale_cuts} we have focused on the minimum scale fitted, $r_\mathrm{min}$. However, we also have to choose the largest scale we fit for, $r_\mathrm{max}$. Furthermore, as our measurement is two-dimensional, we could also impose different scale cuts along or across the line-of-sight. We tested different versions of this analysis, which we summarize here.

We show the impact of changing the largest separation included in our fit, $r_\mathrm{max}$, in Figure \ref{fig:rmax}. We perform the same type of comparison as we did for $r_\mathrm{min}$ in Figure \ref{fig:r_min}. The scales we tested range from $140$\hMpc\ to $200$\hMpc. We find that measurements of \phif\ are unbiased for all the $r_\mathrm{max}$ values we have tried. Furthermore, the improvement in the constraint as we include more data is very weak. Therefore, following eBOSS, we decided to choose $r_\mathrm{max}=180$\hMpc\ in our standard analysis.

\begin{figure*}
    \centering
    \includegraphics[width=1.\textwidth,keepaspectratio]{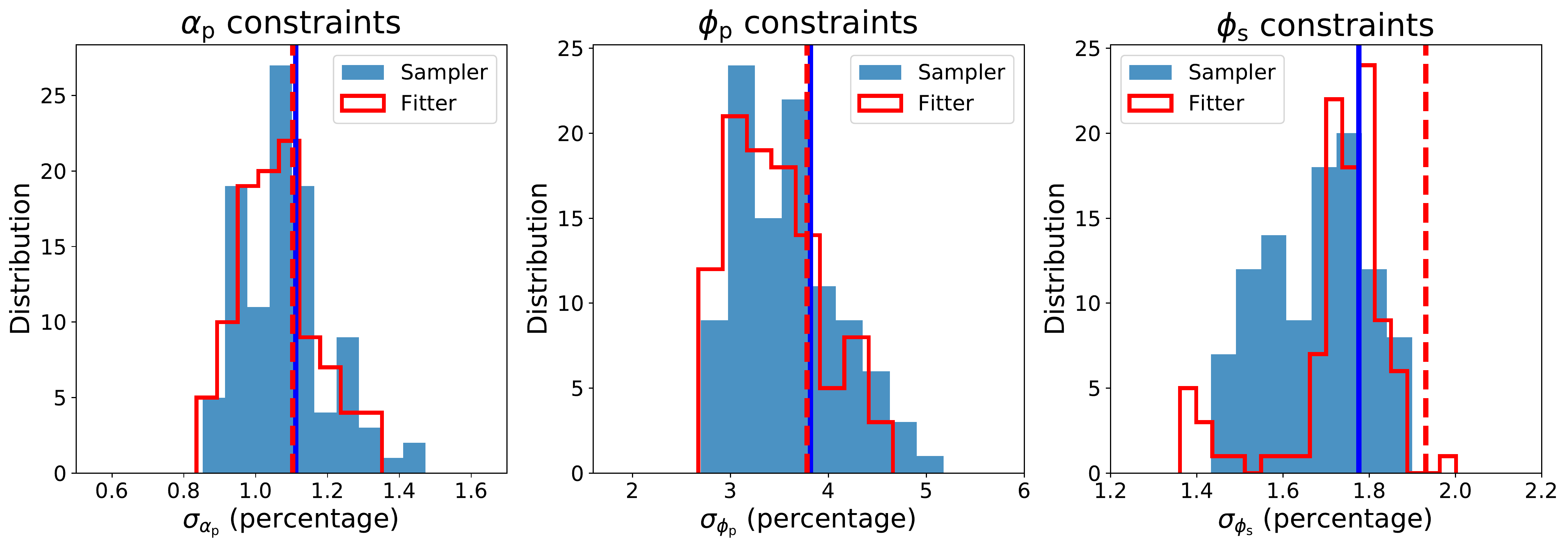}
    \caption{Histograms of the $68\%$ constraints on \alphap, \phip\ and \phis\ when fitting the correlations measured from one hundred uncontaminated eBOSS mocks. We compare results from a fitter which assume a Gaussian posterior, with results from a sampler where we compute the full posterior distributions. The vertical lines represent the standard deviation of the best fitting results for each parameter over the one hundred mocks. We show that when using the sampler, the variance of the population best fits is consistent with the individual constraints for all three parameters.}
    \label{fig:pop_constraints}
\end{figure*}

As most of the contaminants of the \lyaf\ correlation functions have their largest impact along the line-of-sight, we also decided to test a cut in $r_\bot^\mathrm{min}$. This is the smallest transverse scale that we fit, and values $r_\bot^\mathrm{min} > 0$\hMpc\ mean that we are removing the columns in the correlation that lie along the line-of-sight. The systematic error and expected statistical constraint on \phif\ for different values of $r_\bot^\mathrm{min}$ are show in Figure \ref{fig:rtmin}. We find that removing the first correlation column along the line-of-sight ($r_\bot^\mathrm{min}=4$\hMpc) still results in an unbiased measurement. However, removing more data after that starts to introduce a significant systematic bias.\footnote{Note that we are plotting the absolute value of $\Delta\phi$. Therefore, the trend of increasing systematic bias in Figure \ref{fig:rtmin} does not mean the bias is all in the same direction.} This could be explained by the fact that removing the columns along the line-of-sight is degrading our ability to fit the parameters that control the contaminants (e.g. HCDs and metals). Even though these contaminants have their strongest impact along the line-of-sight, they have a small impact on the correlation even at large transverse separations. Therefore, it is still important to model them accurately. Given these results, we chose to not impose any cut in $r_\bot^\mathrm{min}$.

\section{Population constraints}
\label{app:mock_constraints}

As described in \Cref{subsec:mock_pop}, when running the analysis on each of the one hundred mocks, we are restricted to using a minimizer due to computational constraints. This means that the reported constraints on the measured parameters are just estimates due to two reasons. Firstly, they assume the posterior is Gaussian around the best fitting value. Secondly, the best fit for the parameters of interest (\phif, \alphap) is found without properly marginalizing over the nuisance parameters.\footnote{The best fit value of \phis\ reported by the minimizer is conditional on the best fitting values of all the nuisance parameters. However, in the Bayesian framework, the quantity we are interested in is the value of \phis\ conditional on the posterior distributions of the nuisance parameters. Only in the second case is the uncertainty due to these parameters properly taken into account.} While these approximations might be correct for the BAO parameters due to their lack of correlations with the nuisance parameters \citep{Cuceu:2020}, \phis\ definitely has such correlations (Appendix \ref{app:par_corr}). Therefore, in order to study accurate parameter constraints from the population of mocks, we need to use a sampler. 

For the purposes of this appendix, we ran the sampler on correlations computed only using the \lya\ region (no \lyb\ region) from uncontaminated synthetic data sets. These mocks do not include the effects of HCDs, metal absorption or redshift errors \cite[they are also described in][]{duMasdesBourboux:2020}. Therefore, the parameter space is much smaller, allowing for the computation of the full posterior distributions using a sampler. We performed the joint analysis of the auto and cross-correlations for all one hundred uncontaminated mocks.

In Figure \ref{fig:pop_constraints}, we show histograms of the constraints on \alphap, \phip, and \phis\ from the population of one hundred mocks. The main result is the one given by the sampler, in blue. We also plot the standard deviation of the best fitting values for each of the parameters over the population of mocks. This shows that the variance of the parameter values obtained from the one hundred realizations is consistent with the measured uncertainty in individual realizations, which means the uncertainties are properly accounted for.

We also compare results from the sampler with the estimates from the minimizer in Figure \ref{fig:pop_constraints}. While they roughly agree for the BAO parameters (\alphap\ and \phip), this is not the case for \phis. The two give different constraints, as shown by the difference between the red and blue histograms. This is not surprising given the assumptions involved in the fitter constraints. However, the variance of the \phis\ best fits is also different between the two (the red and blue vertical lines). This means that marginalization over the nuisance parameters is important in this case.\footnote{In a frequentist framework, a possible solution for this would be to use these results to compute non-Gaussian $\Delta\chi^2$ values in order to set more accurate confidence intervals on \phis.}

While this has been a rough test due to computational constraints, most of the parameters that \phis\ is correlated with are still present in uncontaminated mocks ($b_{\mathrm{Ly}\alpha}$, $\beta_{\mathrm{Ly}\alpha}$ and $\beta_\mathrm{QSO}$). The main omission is $b_\mathrm{HCD}$. Therefore, it would be interesting to perform this analysis with the effect of HCDs modelled in the mocks. Alternatively, future studies should perform the analysis on the fully contaminated mocks if there is access to enough computational resources.

\section{Parameter correlations}
\label{app:par_corr}

\begin{figure*}
    \centering
    \includegraphics[width=0.7\textwidth,keepaspectratio]{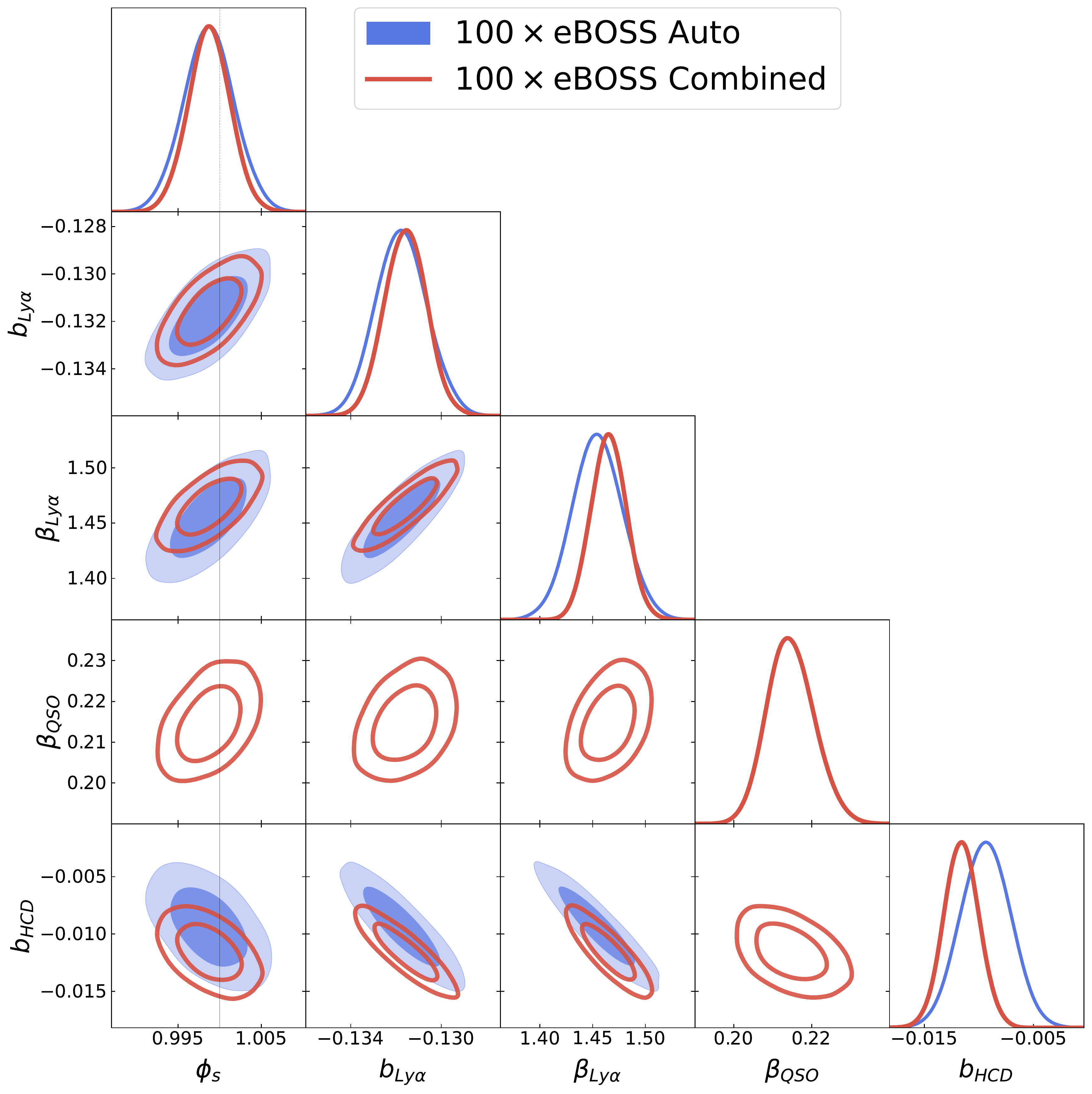}
    \caption{Posterior distributions obtained from fitting the mean of correlation functions computed from one hundred eBOSS mocks. We show results for the joint analysis of the two \lya\ auto-correlations, and for the joint analysis of all four correlations. We only plot the parameters that have correlations with \phis, in order to show its sensitivity to the values of these parameters.}
    \label{fig:kaiser}
\end{figure*}

The advantage of BAO analyses is that they measure a well-defined feature in the correlation function. BAO scale parameters rarely have any significant correlations with the nuisance parameters, which means they are much more robust to different analysis choices when it comes to modelling contaminants. On the other hand, full-shape analyses rely on the ability to accurately model correlation functions over all the scales of interest. We have already shown in Figure \ref{fig:model_tests} what biases are introduced when not modelling certain contaminants. In this appendix, we go into more detail on the sensitivity of \phis\ measurements to other nuisance parameters.

In Figure \ref{fig:kaiser} we show the posterior distributions of the main parameters that are correlated with \phis. The two posteriors are from the fit to the two \lya\ auto-correlations, and from the joint fit of all four correlations. The cross-correlations between \lya\ and quasars cannot be included in such a comparison because the full system of parameters is degenerate.\footnote{See \Cref{subsec:stack_results} above and Section 4 of \cite{Cuceu:2021}} The most important correlations are between \phis\ and the \lya\ bias and RSD parameters, $b_{\mathrm{Ly}\alpha}$ and $\beta_{\mathrm{Ly}\alpha}$. These two are highly correlated between themselves and with the HCD bias, $b_\mathrm{HCD}$. The \lya\ auto-correlation provides most of the information on these parameters. On the other hand, the information on $\beta_\mathrm{QSO}$ comes from the \lya-QSO cross-correlation. However, a measurement of $\beta_{\mathrm{Ly}\alpha}$ (from the auto) is necessary to constrain it due to their degeneracy in the model of the cross-correlation. While we treat $\beta_\mathrm{QSO}$ as a nuisance parameter in this work, future studies could also attempt to extract a measurement of the growth rate of structure from it.


\bsp	
\label{lastpage}
\end{document}